\documentclass[lettersize,journal]{IEEEtran}
\usepackage{amsmath,amsfonts}
\usepackage{algorithmic}
\usepackage{array}
\usepackage[caption=false,font=normalsize,labelfont=sf,textfont=sf]{subfig}
\usepackage{textcomp}
\usepackage{stfloats}
\usepackage{verbatim}
\usepackage{graphicx}
\usepackage{cite}
\usepackage{booktabs}
\usepackage{colortbl}
\usepackage{enumitem} 
\usepackage{makecell}
\usepackage{xcolor}
\usepackage{multirow}
\usepackage{adjustbox}
\usepackage{threeparttable}
\usepackage{tcolorbox}

\usepackage[hidelinks]{hyperref}
\usepackage{cleveref}
\usepackage{xurl}
\makeatletter
\def\UrlAlphabet{%
      \do\a\do\b\do\c\do\d\do\e\do\f\do\g\do\h\do\i\do\j
      \do\k\do\l\do\m\do\n\do\o\do\p\do\q\do\r\do\s\do\t
      \do\u\do\v\do\w\do\x\do\y\do\z\do\A\do\B\do\C\do\D
      \do\E\do\F\do\G\do\H\do\I\do\J\do\K\do\L\do\M\do\N
      \do\O\do\P\do\Q\do\R\do\S\do\T\do\U\do\V\do\W\do\X
      \do\Y\do\Z\do\*\do\-\do\'\do\"\do\_\do\.}
\def\UrlDigits{\do\1\do\2\do\3\do\4\do\5\do\6\do\7\do\8\do\9\do\0}
\g@addto@macro{\UrlBreaks}{\UrlOrds}
\g@addto@macro{\UrlBreaks}{\UrlAlphabet}
\g@addto@macro{\UrlBreaks}{\UrlDigits}
\makeatother

\tcbuselibrary{breakable} 

\def\BibTeX{{\rm B\kern-.05em{\sc i\kern-.025em b}\kern-.08em
    T\kern-.1667em\lower.7ex\hbox{E}\kern-.125emX}}
\usepackage{balance}

\newcommand{\chengwei}[1]{\textcolor{black}{#1}}
\newcommand{\boyuan}[1]{\textcolor{black}{#1}}

\begin{document}

\title{Open Source, Hidden Costs: A Systematic Literature Review on {OSS License Management}}

\author{Boyuan Li\thanks{Boyuan Li and Chengwei Liu contributed equally to this work.}, Chengwei Liu, Lingling Fan, Sen Chen, Zhenlin Zhang, and Zheli Liu
\thanks{Boyuan Li is with DISSec, NDST, College of Computer Science, Nankai University, China. (e-mail: boyuanli@mail.nankai.edu.cn).}
\thanks{Chengwei Liu is with Nanyang Technological University, Singapore (e-mail: chengwei.liu@ntu.edu.sg).}
\thanks{Lingling Fan (Corresponding author), Sen Chen, Zhenlin Zhang, and Zheli Liu are with DISSec, NDST, College of Cryptology and Cyber Science, Nankai University, China (e-mail: linglingfan@nankai.edu.cn, senchen@nankai.edu.cn, liuzheli@nankai.edu.cn).}}

\markboth{Journal of \LaTeX\ Class Files,~Vol.~XX, No.~XX, June~2025}%
{How to Use the IEEEtran \LaTeX \ Templates}

\maketitle

\begin{abstract}
Integrating third-party software components is a common practice in modern software development, offering significant advantages in terms of efficiency and innovation. However, this practice is fraught with risks related to software licensing. A lack of understanding may lead to disputes, which can pose serious legal and operational challenges.
To these ends, both academia and industry have conducted various investigations and proposed solutions and tools to deal with these challenges. However, significant limitations still remain. Moreover, the rapid evolution of open-source software (OSS) licenses, as well as the rapidly incorporated generative software engineering techniques, such as large language models for code (CodeLLMs), are placing greater demands on the systematic management of software license risks. To unveil the severe challenges and explore possible future directions, we conduct the first systematic literature review (SLR) on 80 carefully selected OSS license-related papers, classifying existing research into three key categories, i.e., license identification, license risk assessment, and license risk mitigation. Based on these, we discuss challenges in existing solutions, conclude the opportunities to shed light on future research directions and offer practical recommendations for practitioners. We hope this thorough review will help bridge the gaps between academia and industry and accelerate the ecosystem-wide governance of legitimate software risks within the software engineering community.

\end{abstract}


\section{Introduction}\label{sec:Introduction}
\IEEEPARstart{O}{ver} the past few decades, the rapid development of open-source software (OSS) has been prompted by the spirit of technology sharing, offering significant flexibility and transparency. To avoid reinventing the wheels, an increasing number of developers choose to incorporate open-source artifacts into their projects across various levels of granularity to reduce development efforts. 
For instance, as reported by Census II, OSS constitutes 70-90\% of any given piece of modern software solutions~\cite{ASummary30:online}. Approximately 70\% of the code on GitHub actually consists of code clones from existing repositories~\cite{cloneOnGithub}, or even code snippets posted on Stack Overflow~\cite{cloneOnSO}.


Every coin has two sides. The freedom of OSS does not negate the need for respecting the discipline in use. 
While providing convenience to developers, OSS also introduces various legal risks in practical applications. 
To this end, software licenses have been widely adopted to regulate the disciplines and protect the copyright of software owners, and specifically, OSS licenses are proposed to balance copyright protection.
Downstream users must adhere to the obligations of software licenses before enjoying the rights they are granted.

However, due to the complexity of software licenses and diverse ways of software reuse, it is still non-trivial to identify the potential legitimate risks when reusing OSS, and disputes over license violations during the development process are {also common.}
{Many big enterprises, such as Google, TikTok, and Microsoft 
have faced lawsuits and controversies related to software license violations.}
For instance, {Google was found to have copied part of Java SE API code for its Android platform without a granted license~\cite{googleOracle}, TikTok Live Studio was accused of violating GPL by incorporating code from OBS Studio without adhering to requirements~\cite{tiktokOBS}, {{GitHub Copilot} is also facing a class-action lawsuit for generating code without proper attribution to human-written code (e.g., name and copyright notice)}, potentially infringing on copyrights and violating various OSS licenses~\cite{copilotSue, copilotSue2}.



To properly identify, manage, and mitigate the potential threats of these legitimate risks, both academia and industry have been devoted to extensively investigating the license risks in the open-source environment. Many related research works have been carried out, such as conducting empirical studies~\cite{intetrationPatterns, baolingfenglicense}, developing automated tools~\cite{ninka, fossology}, employing reviews and comparative analyses~\cite{license_tool_metrics, tuunanen2021tool}, from various aspects, including but not limited to license incompatibilities~\cite{SPDXcompatibility, lidetector, pypi}, conflicts~\cite{DIKE}, inconsistencies~\cite{inconsistenciesInLarge-scaleProjects, inconsistencyInLargeCollections}, violations~\cite{SPDX1, OSSPolice, dockerCompatibility}, and compliance~\cite{CBDG, Kenen}, to facilitate better understandings of software licenses.
However, despite extensive research efforts, the rapid development of software products and their corresponding licenses has introduced significant challenges in understanding licensing~\cite{doDevelopersUnerstandLicenses?, developerRef1}, which is evidenced by the sparse application of research tools in industrial settings~\cite{wintersgill2024law}. 
{In parallel with academia’s efforts, the industry has also been devoted to the standardization of open-source compliance practices to address the complexity of software licensing. For instance, the OpenChain Project~\cite{coughlan2020standardizing} has provided structured guidelines and best practices to help organizations build robust open-source compliance programs~\cite{iso5230}. The Software Package Data Exchange (SPDX) specification established a standardized framework for the structured presentation of license and copyright information~\cite{stewart2010software}, which serves as a foundation of software bills of materials (SBOMs)~\cite{ntia_sbom} and provides comprehensive transparency regarding software components and their associated licensing details. 
Based on these, Software Composition Analysis (SCA) tools are further developed and incorporated to facilitate compliance by automating the identification of licenses and potential conflicts~\cite{SCA_Sonatype,SCA_BlackDuck, SCA_Mendio, openMLWhitepaper}.} 
However, these efforts are still undermined by the poor understanding and implementation of software license restrictions~\cite{wintersgill2024law}. 

To the best of our knowledge, no existing work has systematically reviewed and studied the workflow, as well as the capability of state-of-the-art (SOTA) approaches, of the current practice of OSS license management. {In this paper, to bridge this gap,} we collect research on license management and, guided by the functionalities of industry tools, propose a taxonomy that categorizes these works into three major objectives: license identification, license risk assessment, and license risk mitigation. Based on this, we conduct the first systematic literature review to delve into the management process of OSS licenses, based on which, we aim to unveil the persistent challenges and shed light on future research directions, for better ecosystem-wide governance of OSS licenses. 


Our main contributions to this paper are as follows:

\begin{itemize}
  \item[$\bullet$] \textbf{A comprehensive collection and review of OSS license management related research.} Based on the well-defined SLR method, we conduct a thorough analysis of 80 well-collected papers related to software license management.
  
  \item[$\bullet$] \textbf{{A taxonomy of OSS license management research based on industrial practice.}}
  {We propose a taxonomy of OSS license management research guided by industrial tool functionalities and workflows, including license identification, license risk assessment, and license risk mitigation. This taxonomy reflects the alignment between academic research and industrial practice.} 
  
  
  \item[$\bullet$] \textbf{{A discussion on challenges, opportunities, and recommendations for different stakeholders.}} We extend a thorough discussion on existing challenges and limitations of existing license management, based on which we summarize the opportunities to propose possible future directions for license management. {We also provide practical license management recommendations based on the current state of research for different stakeholders.}
\end{itemize}

The remainder of this paper is organized as follows:
{Section~\ref{sec:Concept} defines several important concepts and illustrates them with examples.
Section~\ref{sec:collect} outlines the research methodology for paper collection and taxonomy construction. Section~\ref{sec:review} presents a comprehensive review of the selected studies, which serves as the basis for the in-depth discussion of key challenges, research opportunities, practical recommendations, and threats to validity in Section~\ref{sec:discussion}.}

\section{Background}\label{sec:Concept}

{As legal agreements that outline the conditions for the use, modification, and distribution of software, \textbf{OSS licenses} are commonly used to define the responsibilities and obligations of users, and provide legal protection for the copyright of developers to prevent their open-source software from being illegally shared and reused~\cite{licenseDefination}. However, considering the continuous evolution of OSS licenses, their application, and terminologies in the OSS ecosystem, we first clarify some key concepts in this paper to avoid misunderstanding.}

{\noindent $\bullet$ \textbf{License Proliferation}: It describes the phenomenon of the increasingly complicated legal environment for developers, companies, and users caused by the increasing complexity and prevalence of distinctive software licenses~\cite{proliferationDefination}. This occurs as licenses are customized to meet specific legal, ethical, or technological needs.
{Black Duck noted in their report that, among the codebases they audited in 2022 and 2023, 31\% didn't contain standard licenses~\cite{synopsys2024report}.}
While these varied licenses provide tailored terms for software use and distribution, they also make compliance and interoperability more difficult, leading to confusion, higher legal risks, and potential misuse of software.}

{\noindent $\bullet$ \textbf{License Compatibility}:  It refers to the ability of two software licenses to coexist without violating the terms of either~\cite{WikiDefinationCompatibility}. This compatibility is essential for the legal reuse of software, especially when components or libraries under different licenses are combined in a single project. Understanding and ensuring license compatibility helps developers avoid legal pitfalls and ensures that software remains free and open for use and redistribution under specified terms.}

{\noindent $\bullet$ \textbf{License Non-Compliance}: License non-compliance in OSS involves failing to meet the specific conditions set by OSS licenses, such as proper attribution, releasing modified source code, or including the original license upon redistribution~\cite{complianceDefination}. It often results from a lack of awareness or understanding of licensing terms, inadequate management of software assets, or the complexities involved in integrating multiple software components with different licensing requirements. Addressing license non-compliance is crucial for maintaining the legal and ethical integrity of software development and usage.}

{\noindent $\bullet$ \textbf{License Term}: License term (a.k.a. license clause~\cite{gangadharan2012Managing,ALP}) refers to rights granted and obligations imposed on users when using software~\cite{lidetector, catchTheButterfly}. For example, the license term ``redistribution" might appear as ``distribute copies" or ``redistribute" in licenses. Compared to license types, license terms provide a more granular framework for management.}

{\noindent $\bullet$ \textbf{License Exception}:  License exceptions create exemptions from specific license conditions or extend permissions beyond the original terms~\cite{exceptionDefination}. A well-known example is the GPL linking exception~\cite{GPLLinkingException}, which allows programs to link to GPL-licensed libraries without fully complying with all GPL terms. Such exceptions function similarly to customized licenses but help mitigate license proliferation.}

\section{{Research Methodology}}\label{sec:collect}


{To further understand the latest developments in OSS license management, we conducted an SLR on the subject and followed existing well-established guidelines~\cite{SLR_guidelines, SLR_snowballing, LiteratureReview_StaticAndroid, LiteratureReview_3rdLibraryAndroid}.} {\Cref{SLRprocess}, adapted from the PRISMA flow diagram template~\cite{page2021prisma}, illustrates the complete workflow for paper retrieval and selection.}

    \begin{figure}
      \centering
      \includegraphics[width=\linewidth]{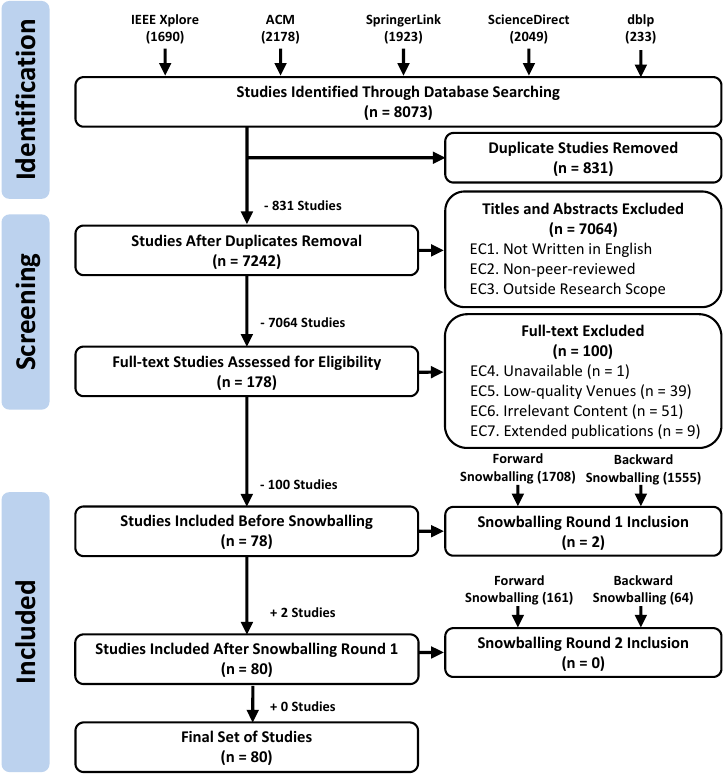}
      \caption{{Overview of literature retrieval and selection process}}
      \label{SLRprocess}
    \end{figure}

\subsection{Search Strategy}
{To ensure the systematic and comprehensive collection of papers, we developed the following search strategies:}
    
    \subsubsection{\textbf{Search Scope}} 
    {Our study aims to provide a comprehensive review of academic research on how developers manage project licenses. To this end, we seek to identify all relevant papers that focus on OSS license management as their primary research objective, such as selecting, tracking, altering, or complying with OSS licenses.}

    \subsubsection{\textbf{Search Keywords}}
    After clarifying the search scope, we defined search keywords aiming at fully covering all relevant research papers within the search scope. 
    The final keywords are classified into two groups as
    illustrated in~\Cref{tab:keyword}.
    $\mathit{Group~1}$ limits the research coverage and consists of ``oss" and its synonyms. Since the term ``license" itself has various meanings and is widely used in industry and different fields, we incorporated a set of restrictive modifiers related to OSS to filter out irrelevant research initially. {However, to avoid being overly restrictive, we selected inherently broad terms like ``software," ``open source," and {``software supply chain{,}"} which allowed us to cast a wide net and later refine the results through manual review.} 
    {$\mathit{Group~2}$ defines the focus of our research and consists of ``license" and its variants, as we consider the presence of the term ``license" to be a fundamental criterion for determining the relevance of a paper.} 
    To ensure the completeness of the collected papers, we included the wildcard character ``*" in our search keywords.   
    Finally, the search term is defined by the expression $\mathit{g1}$ \textbf{AND} $\mathit{g2}$, where $\mathit{g1} \in Group\mathit{~1}$ and $\mathit{g2} \in Group\mathit{~2}$.

\begin{table}[t]
\centering
\caption{Search Keywords}
\label{tab:keyword}
\begin{tabular}{cl}
\toprule
\textbf{Group} & \textbf{Keywords} \\
\midrule
    1 & floss; foss; oss; ``open source"; software:\\
      & ``open source software"; ``software supply chain"\\
    2 & licens*: license; licenses; licensed; licensing\\
\bottomrule
\end{tabular}
\end{table}

    \subsubsection{\textbf{Search Database}}
    Our search was conducted across five well-known digital databases: ACM Digital Library~\cite{acm}, IEEE Xplore Digital Library~\cite{IEEELibrary}, SpringerLink~\cite{SpringerLink}, ScienceDirect~\cite{ScienceDirect} and {DBLP}~\cite{dblp}. 
    It's worth noting that different repositories and search engines utilize unique syntax rules and search fields for advanced querying. For instance, the ACM Digital Library allows the use of wildcards and supports keyword searches across abstracts, while DBLP does not support wildcards and limits searches to the paper title field. We thus tailored our search strategies to the specific rules of each database. The publication dates of the papers were limited to January 2000 through September 2024, aligning with the timing of the search. There were duplicates among the papers retrieved from different databases, so we merged the results to obtain a list of unique papers.}
   {After this, we collected 7,242 papers in total as the initial results.}

    \subsection{Study Selection}\label{subsec:sca}
    \subsubsection{\textbf{Paper Exclusion}}
    The papers obtained by keyword-based searching may introduce a significant volume of papers beyond our research scope. To refine the results, we implemented a set of exclusion criteria as below to filter out papers:

    \noindent \textbf{EC1. Papers not written in English.} Since English is the predominant language for international scholarly communication, English research can cover the widest range of innovative research and academic perspectives, thereby commanding greater acceptance and influence.
        
    \noindent \textbf{{EC2. Non-peer-reviewed publications.}} {We excluded papers that had not undergone rigorous peer review (e.g.{,} theses, papers from arXiv, or unpublished manuscripts).}
        
    \noindent \textbf{{EC3. Out-of-scope papers.}} {We excluded papers unrelated to our research scope (e.g., those concerning proprietary licenses or dataset construction), as well as license-related work in other fields or industries (e.g., driver's licenses).} 

    {We first filter the collected papers by examining their titles and abstracts against EC1$\sim$EC3 and obtain 178 papers. For each paper, we downloaded the full texts and manually went through their content to further exclude irrelevant papers. Specifically, we adapted four additional exclusion criteria during the manual examination:}

    
    \noindent \textbf{{EC4. Studies whose full-text is unavailable.}} {We ignored papers that we can only find an abstract rather than full text from all available sources (including the scholar databases and Google Scholar~\cite{googleScholar}), to avoid misunderstanding with only titles and abstracts.}
    


    \noindent \textbf{{EC5. Papers published in low-quality venues.}} {Following the existing top-venue SLR~\cite{croft2022data}, we excluded papers published in venues that were not included in the latest CORE ranking list to avoid insufficiently validated results~\cite{coreRanking1,coreRanking2}.}

    \noindent \textbf{{EC6. Papers with limited relevance to the research scope.}} {Some papers were also found with limited relevance to our scope after going through the full text, so we also exclude them for better concentration. For instance, a paper may mention license compliance as motivation in its summary, but it does not actually address license management in the full text.}

    \noindent \textbf{{EC7. Extended publications.}} {We retained only the most extended version of a work. For instance, if a study was first presented as a conference paper and later extended into a journal article, we only keep the latter one with the most complete content.}

 
    {{In this step, a total of 100 papers were excluded: 1 for unavailability (EC4), 39 from low-quality venues (EC5), 51 due to irrelevant content (EC6), and 9 as extended publications (EC7). As a result, 78 papers were retained as our preliminary selection.
    The full-text screening was carried out by a three-member team comprising one doctoral candidate with two years of OSS-license research experience and two senior researchers, each with more than five years in the field. All three reviewers independently reviewed each paper to decide on its inclusion. As a result, inclusion disagreements occurred in only five of the 78 reviewed papers, and the average Cohen’s $\kappa$ was 0.95, indicating a high level of agreement. The disagreements were finally resolved through discussion and consensus among the authors.}}

    \subsubsection{\textbf{Snowball Sampling}}

    {To ensure no relevant research works were overlooked, we further conducted snowball sampling on the 78 collected papers. Specifically, for each round of snowball sampling, we recursively identified relevant papers that 1) cite it or 2) were cited by it, through Google Scholar~\cite{googleScholar}, for each collected paper, until no more papers were included.
    {Specifically, in the first round, we applied forward and backward snowballing to the 78 seed papers, retrieving 1,708 forward citations and 1,555 backward references. After merging the two sets and removing duplicates, 3,091 unique candidate papers remained. These candidates were then assessed against EC1$\sim$EC7. 70 papers met all criteria. Of these, 68 were already included in the seed set, so only two papers remained~\cite{codeSiblings,ghapanchi2011impact}. A second round of snowballing based on these two papers returned no further eligible studies. Consequently, snowballing contributed two papers, bringing the total number of primary studies in our SLR to 80.}}

        \begin{figure}
          \centering
          \includegraphics[width=0.48\textwidth]{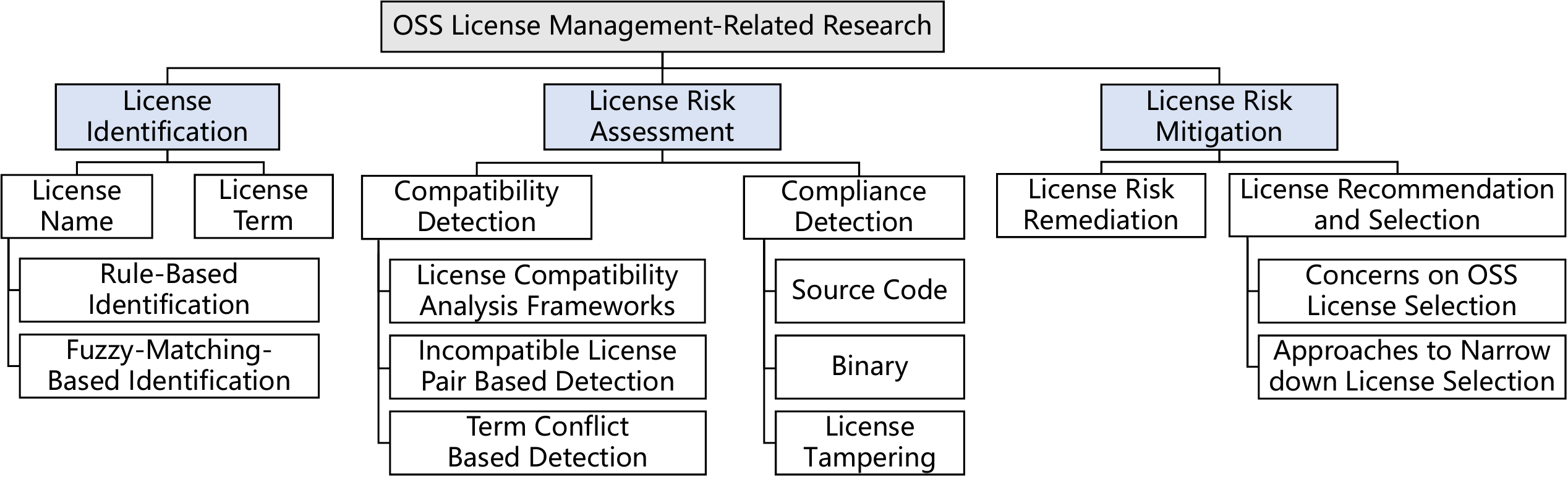}
          \caption{The Overview of research works on OSS license management}
          \label{taxonomy}
        \end{figure} 
        

\begin{table*}[]
\centering
\caption{{Overview of functionalities of industrial license management tools}}
\label{tab:SCAtoolFunctions}

\resizebox{0.9\textwidth}{!}{

\begin{tabular}{l|c|cc|ccc|c}
\Xhline{0.8px}
\multicolumn{1}{c|}{}                                     &                                                                                             & \multicolumn{2}{c|}{\textbf{License Risk Assessment}}                       & \multicolumn{3}{c|}{\textbf{License Risk Mitigation}}                                                                                                                                                                                               &                                     \\ \cline{3-7}
\multicolumn{1}{c|}{\multirow{-2}{*}{\textbf{{Industrial} Tools}}} & \multirow{-2}{*}{\textbf{\begin{tabular}[c]{@{}c@{}}License\\ Identification\end{tabular}}} & \multicolumn{1}{c|}{\textbf{Compatibility}}         & \textbf{Risk Profile} & \multicolumn{1}{c|}{\textbf{\begin{tabular}[c]{@{}c@{}}Attribution\\ Notice\end{tabular}}} & \multicolumn{1}{c|}{\textbf{\begin{tabular}[c]{@{}c@{}}Remediation\\ Options\end{tabular}}} & \textbf{\begin{tabular}[c]{@{}c@{}}License\\ Selection\end{tabular}} & \multirow{-2}{*}{\textbf{\begin{tabular}[c]{@{}c@{}}Publicly Accessible\\Information Sources\end{tabular}}} \\ \hline
\rowcolor[HTML]{E6E6E6} 
Sonatype~\cite{SCA_Sonatype}                                                  & \textcolor[HTML]{00A551}{\checkmark}                                                                                      & \multicolumn{1}{c|}{\cellcolor[HTML]{E6E6E6}}       & \textcolor[HTML]{00A551}{\checkmark}                & \multicolumn{1}{c|}{\cellcolor[HTML]{E6E6E6}\textcolor[HTML]{00A551}{\checkmark}}                                        & \multicolumn{1}{c|}{\cellcolor[HTML]{E6E6E6}\textcolor[HTML]{00A551}{\checkmark}}                                              & \multicolumn{1}{c|}{\cellcolor[HTML]{E6E6E6}}                                                                   &  \cite{SCA_Sonatype2, SCA_Sonatype3, SCA_Sonatype4}                                   \\
Black Duck~\cite{SCA_BlackDuck}                                                & \textcolor[HTML]{00A551}{\checkmark}                                                                                      & \multicolumn{1}{c|}{\textcolor[HTML]{00A551}{\checkmark}}                         & \textcolor[HTML]{00A551}{\checkmark}                & \multicolumn{1}{c|}{\textcolor[HTML]{00A551}{\checkmark}}                                                                      & \multicolumn{1}{c|}{\textcolor[HTML]{00A551}{\checkmark}}                                                                &                                                                      &  \cite{SCA_BlackDuck2, SCA_BlackDuck4, SCA_BlackDuck5, SCA_BlackDuck7,SCA_BlackDuck9}                                   \\
\rowcolor[HTML]{E6E6E6} 
Mend.io~\cite{SCA_Mendio}                                                   & \textcolor[HTML]{00A551}{\checkmark}                                                                                      & \multicolumn{1}{c|}{\cellcolor[HTML]{E6E6E6}\textcolor[HTML]{00A551}{\checkmark}} & \textcolor[HTML]{00A551}{\checkmark}                & \multicolumn{1}{c|}{\cellcolor[HTML]{E6E6E6}\textcolor[HTML]{00A551}{\checkmark}}                                        & \multicolumn{1}{c|}{\cellcolor[HTML]{E6E6E6}}                                              &                                                                      & \cite{SCA_Mendio2, SCA_Mendio3, SCA_Mendio4, SCA_Mendio5, SCA_Mendio6, SCA_Mendio7}                                    \\
Veracode~\cite{SCA_Veracode}                                                  & \textcolor[HTML]{00A551}{\checkmark}                                                                                      & \multicolumn{1}{c|}{}                               & \textcolor[HTML]{00A551}{\checkmark}                & \multicolumn{1}{c|}{}                                                                      & \multicolumn{1}{c|}{\textcolor[HTML]{00A551}{\checkmark}}                                                                      &                                                                      & \cite{SCA_Veracode2,SCA_Veracode3}                                    \\
\rowcolor[HTML]{E6E6E6} 
Snyk~\cite{SCA_Snyk}                                                      & \textcolor[HTML]{00A551}{\checkmark}                                                                                      & \multicolumn{1}{c|}{\cellcolor[HTML]{E6E6E6}}       & \textcolor[HTML]{00A551}{\checkmark}                & \multicolumn{1}{c|}{\cellcolor[HTML]{E6E6E6}}                                              & \multicolumn{1}{c|}{\cellcolor[HTML]{E6E6E6}}                                              &                                                                      & \cite{SCA_Snyk2, SCA_Snyk3}                                    \\
Checkmarx~\cite{SCA_Checkmarx}                                                 & \textcolor[HTML]{00A551}{\checkmark}                                                                                      & \multicolumn{1}{c|}{}                         & \textcolor[HTML]{00A551}{\checkmark}                & \multicolumn{1}{c|}{}                                                                      & \multicolumn{1}{c|}{\textcolor[HTML]{00A551}{\checkmark}}                                                                &                                                                      &  \cite{SCA_Checkmarx2,SCA_Checkmarx3, SCA_Checkmarx4}                                   \\
\rowcolor[HTML]{E6E6E6} 
Revenera~\cite{SCA_Revenera}                                                  & \textcolor[HTML]{00A551}{\checkmark}                                                                                      & \multicolumn{1}{c|}{\cellcolor[HTML]{E6E6E6}}       & \textcolor[HTML]{00A551}{\checkmark}                & \multicolumn{1}{c|}{\cellcolor[HTML]{E6E6E6}\textcolor[HTML]{00A551}{\checkmark}}                                        & \multicolumn{1}{c|}{\cellcolor[HTML]{E6E6E6}\textcolor[HTML]{00A551}{\checkmark}}                                              &                                                                      & \cite{SCA_Revenera2, SCA_Revenera3, SCA_Revenera4}                                    \\
JFrog~\cite{SCA_JFrog}                                                     & \textcolor[HTML]{00A551}{\checkmark}                                                                                      & \multicolumn{1}{c|}{}                               &                 & \multicolumn{1}{c|}{}                                                                & \multicolumn{1}{c|}{}                                                                      & \textcolor[HTML]{00A551}{\checkmark}                                                                     &  \cite{SCA_JFrog2, SCA_JFrog3}                                   \\
\rowcolor[HTML]{E6E6E6} 
Palo Alto Networks~\cite{SCA_Palo}                                        & \textcolor[HTML]{00A551}{\checkmark}                                                                                      & \multicolumn{1}{c|}{\cellcolor[HTML]{E6E6E6}}       & \textcolor[HTML]{00A551}{\checkmark}                & \multicolumn{1}{c|}{\cellcolor[HTML]{E6E6E6}}                                        & \multicolumn{1}{c|}{\cellcolor[HTML]{E6E6E6}\textcolor[HTML]{00A551}{\checkmark}}                                        &                                                                      &  \cite{SCA_Palo2}                                   \\
GitHub~\cite{SCA_GitHub}                                                    & \textcolor[HTML]{00A551}{\checkmark}                                                                                      & \multicolumn{1}{c|}{}                               &                       & \multicolumn{1}{c|}{}                                                                      & \multicolumn{1}{c|}{}                                                                      & \textcolor[HTML]{00A551}{\checkmark}                                                               & \cite{SCA_GitHub2, SCA_GitHub3}                                      \\
\rowcolor[HTML]{E6E6E6} 
GitLab~\cite{SCA_GitLab}                                                    & \textcolor[HTML]{00A551}{\checkmark}                                                                                      & \multicolumn{1}{c|}{\cellcolor[HTML]{E6E6E6}\textcolor[HTML]{00A551}{\checkmark}} & \textcolor[HTML]{00A551}{\checkmark}                & \multicolumn{1}{c|}{\cellcolor[HTML]{E6E6E6}}                                              & \multicolumn{1}{c|}{\cellcolor[HTML]{E6E6E6}}                                              &                                                                      &  \cite{SCA_GitLab2, SCA_GitLab3}                                   \\
Aqua \boyuan{Security}~\cite{SCA_Aqua}                                             & \textcolor[HTML]{00A551}{\checkmark}                                                                                      & \multicolumn{1}{c|}{\textcolor[HTML]{00A551}{\checkmark}}                         & \multicolumn{1}{c|}{\textcolor[HTML]{00A551}{\checkmark}}                       & \multicolumn{1}{c|}{}                                                                      & \multicolumn{1}{c|}{}                                                                      &                                                                      &     \cite{SCA_Aqua2, SCA_Aqua3}                                \\ \Xhline{0.8px}
\end{tabular}

}

\end{table*}

    \subsection{Paper Classification}
    
{We aim to develop a classification framework for the collected papers. However, the academic community has not yet established a widely accepted taxonomy in this area. Therefore, we draw on established industry practices, which offer mature workflows and widely adopted tools, to inform the structure of our taxonomy.
To explore how industry solutions address OSS license issues, we focused on SCA tools, which are the most commonly used technologies for license management in practice. Guided by the Forrester Wave evaluation report~\cite{forresterSCA}, a well-established and influential industry evaluation, we selected 12 leading SCA products with license-related capabilities. We then analyzed their publicly available resources, such as introductory materials, whitepapers, datasheets, and technical documentation, to identify the license-related functionalities implemented in these products.} 

{The license-related capabilities of these tools are summarized in~\Cref{tab:SCAtoolFunctions}. These functionalities encompass three key directions: license identification, license risk assessment, and license risk mitigation.}
{Specifically, tools start by inspecting the source code and configuration files of user projects, and identify components and code snippets that are possibly third-party libraries. Next, they map the features (i.e., names, version numbers, code hashes, etc.) and query pre-constructed feature databases to confirm the suspicious third-party artifacts and relate them to known security issues or licenses. Then, they assess potential risks, and follow the SBOM standard to generate reports for users. Finally, some tools can provide suggestions or solutions for users to remediate identified issues.}



{We observed that the papers could also map onto these functional areas. However, the coarse groups failed to capture the nuanced differences among works.
To this end, we first extracted the research objective of each paper, such as extracting license terms or assessing license compatibility. These objectives were then integrated into a twelve-category taxonomy shown in~\Cref{taxonomy}. Three authors independently judged whether each paper belonged to specific categories, resulting in 960 classification records. Across all author pairs, there were a total of 66 disagreements, corresponding to classification differences in 21 unique papers. The average Cohen’s $\kappa$ was 0.87, indicating substantial agreement. Disagreements were resolved through discussion and consensus. The complete classification is available at~\cite{list}.}



\section{{Literature Review}}\label{sec:review}


{In this section, we discuss the existing works on OSS license management by the most concerned functionalities, i.e., license identification, license risk assessment, and license risk mitigation. For each functionality, we discuss the existing works by their different objectives, granularity, and approaches in detail, as presented in~\Cref{taxonomy}, and we also explore their relevance within the context of current industry perspectives.}

    \subsection{License Identification}

        The gradual excessive reuse of software components in modern software development has fostered various needs for adopting software licenses. Consequently, various software licenses have been proposed to fit the newly emerged scenarios of software reuse, leading to the challenging identification of software licenses. Generally, the objectives of license identification experienced a migration from license names to detailed license terms, as well as changes in adopted techniques. \Cref{tab:identification_research} presents the timeline of existing research works.


        \subsubsection{\textbf{License Name Identification}}

        {The primary purpose of license name identification is to determine which license governs a given software scope, such as repositories, folders, files, or code snippets, to facilitate follow-up risk assessment. 
        Existing research on identifying OSS license names can be divided into two major strategies: 
        1) Rule-based identification, which relies on predefined rules to capture identical information from license text to distinguish different OSS licenses,
        and 2) Fuzzy-matching-based identification, which directly distinguishes OSS licenses by their textual similarity with known licenses. 
        In this section, we introduce the development of these two main strategies in detail.}
        


        \noindent $\bullet$ \textbf{Rule-Based Identification}: 
        {ASLA}~\cite{ASLA} and {FOSSology}~\cite{fossology} are the first to introduce the capability of identifying software licenses by adopting regular expression templates to match the textual declaration of software licenses. 
        {As the pioneer prototype, although ASLA only supports the identification of 37 different types of software licenses, it proposed an extensive framework that allows for the expansion of regular expression templates.}
        Similarly, {Nomos}, the license identification tool integrated into {FOSSology}, employs short seed expressions to extract areas of interest, standardizes these areas, and scans larger text sections to map to licenses. Until we drafted this paper, the rule-based templates in {Nomos}~\cite{nomos} supported the identification of over 1,700 licenses.
        
        However, 
        {as pointed out by German et al.~\cite{ninka},}
        these tools suffer from potential inaccuracy due to various scenarios of license declaration, such as mixing with source code, replacing with URL links, intended customization, and condition altering. 
        To this end, {Ninka}~\cite{ninka} addressed these challenges with a sentence-based matching algorithm, which utilizes filtering keywords, equivalence phrases, sentence-token expressions, and license rules, to strengthen the accuracy and robustness of rule-based matching for 112 popular licenses. 
        {To further enhance the scalability of Ninka, especially when new licenses are proposed, Higashi et al.~\cite{clusterninka2, clusterninka3} proposed automated solutions to cluster license statements from unknown licenses and extract sequential patterns to facilitate the automated generation of license matching rules.}

        {Apart from license text, license names are also embedded with rich information for identification. Especially, relying on license names logged in the metadata of third-party libraries is most commonly used to retrieve license information by SCA tools. To this end, Wu et al.~\cite{baolingfenglicense} proposed an extraction and algorithm based on AC (Aho-Corasick) automaton~\cite{10.1145/360825.360855} and regular expression matching to precisely identify the standardized licenses referred by variations of license names.}

        \begin{table}[t]
        \centering
        \caption{A Summary of License Identification Research}
        \label{tab:identification_research}
        \resizebox{0.49\textwidth}{!}{
        \begin{tabular}{c|clr}
        \Xhline{0.8px}
        \textbf{Granularity} & \textbf{Method} & \textbf{Tool or First Author} & \textbf{Year}\\
        \hline
        \multirow{11}{*}{\textbf{License Name}} &\multirow{6}{*}{\makecell[c]{Rule-Based\\Identification}} 
        & Nomos~\cite{fossology}  & 2008\\
         &  & ASLA~\cite{ASLA}  & 2009\\
         &  & Ninka~\cite{ninka}   & 2010\\
         &  & Higashi et al.~\cite{clusterninka2}   & 2019\\
         &  & Higashi et al.~\cite{clusterninka3}   & 2023\\
         &  & Wu et al.~\cite{baolingfenglicense}   & 2024\\
        \Xcline{2-4}{0.5px}
         & \multirow{5}{*}{\makecell[c]{Fuzzy-Matching\\-Based Identification}} 
         &  Di Penta et al.~\cite{jarArchives}   & 2010\\
         &  & LChecker~\cite{LChecker} & 2010\\
         &  & Monk~\cite{fossology} & 2015\\
         &  & SORREL~\cite{Sorrel}   & 2021\\
         &  & Wolter et al.~\cite{inconsistenciesOnGithub} & 2023\\

        \hline
        
        \multirow{6}{*}{\textbf{License Term}}
         & \multirow{6}{*}{-} & Vendome et al.~\cite{machineLearningExcepions}   & 2017\\
         & & FOSS-LTE~\cite{fosslte}   & 2017\\
         & & DIKE~\cite{DIKE}   & 2023\\
         & & LiDetector~\cite{lidetector}   & 2023\\
         & & LiSum~\cite{lisum}   & 2023\\
         & & LiResolver~\cite{LiResolver}   & 2023\\
        \Xhline{0.8px}
        \end{tabular}
        }
        \end{table}

        \noindent $\bullet$ \textbf{Fuzzy-Matching-Based Identification}: 
        {Although rule-based approaches have been demonstrated to be effective at license identification, their accuracy highly relies on the quality of pre-defined rules. It is challenging to define rules to perfectly fit all new-emerging licenses, especially under the context of license proliferation. 
        {For example, as revealed by Higashi et al.~\cite{clusteringNinka}, the strict regular expression patterns often misclassified licenses when confronted with minor spelling errors or variant notations.}
        In this context, fuzzy matching methods were introduced, utilizing text similarity techniques and machine learning classifiers to perform generalized matching across the license text.}
        
        For instance, {FOSSology} 3 integrated a new license identification tool, {Monk}~\cite{fossology,monk} to empower fuzzy license identification by matching licenses by text similarity. In this case, users can easily support new licenses by simply including their textual content for similarity analysis.
        %
        Similarly, {SORREL}~\cite{Sorrel}, a license management plug-in for IntelliJ IDEA, proposes a hybrid approach for identifying 16 popular licenses in Java projects, by introducing both machine learning classifier and Sørensen-Dice similarity coefficient, to further facilitate the detection of license incompatibilities. 
%

        Some researchers also rely on the matching of source code to trace the original OSS licenses. 
        Di Penta et al.~\cite{jarArchives} conducted fuzzy-matching to the source code of jar archives. They extracted textual information and bytecode of classes and queried the search engine to retrieve similar code, as well as their licenses. 
        %
        {Zhang et al.~\cite{LChecker} utilizes Google Code Search service to check whether a local file exists in an OSS project, and therefore, identify its original licenses.}

        {Moreover, licenses may also be declared in different entities within a project{, ranging from source-code comments, metadata files, to dedicated files such as \textit{LICENSE}, \textit{README}, or \textit{COPYING}.
        Although such dispersion does not, in itself, hinder identification tools from extracting licenses, which typically rely on pattern-matching heuristics and are therefore indifferent to the specific textual format, it still brings difficulties.}
        Wolter et al.~\cite{inconsistenciesOnGithub} found that many projects exhibit inconsistencies in license declarations for the same component across these entities. It highlights the limitations of current tools, which struggle to reconcile licenses across diverse project entities accurately.
        }



        \subsubsection{\textbf{License Term Identification}}

        

        {The existing study~\cite{stackoverflowKapitsaki} has shown that developers often have significant doubts and misunderstandings about existing licenses, highlighting a disconnect between their needs and the current licensing landscape. 
        As a result, a multitude of new licenses have been created to address concerns raised by different stakeholders, leading to license proliferation.}
        {In particular, the prevalence of license customization~\cite{nonOsiApprovedLicenses, lidetector}, challenges not only the identification of OSS licenses, but also the risk assessment after identifying the various OSS licenses. To this end, some researchers turned to inspect the detailed license statements to identify license terms and assess possible risks on rights and obligations directly, regardless of which license they are.}

        As the fundamental of term-based conflict analysis, most existing works construct their own models of license terms and map text to these terms to facilitate the identification of popular licenses. {FOSS-LTE}~\cite{fosslte} initially manually summarized 37 types of license term templates and utilized Latent Dirichlet Allocation (LDA) for topic modeling to discern abstract topics within the license texts. These topics are then matched to the extracted license terms using vector similarity measures, ensuring that each term is associated with the most representative topics. 
        However, {as discussed in Cui et al.~\cite{DIKE}}, the pre-defined templates should incorporate complex terms with specific attitudes, which require lots of manual effort and could result in low coverage of license types.
        To solve this, {they proposed {DIKE}}, which focused on the identification of 12 common license terms with inferences of the attitude for each license term, offering greater flexibility in identifying more licenses than FOSS-LTE. By modeling 3,256 licenses, DIKE constructed a \textit{License-Terms-Responsibility} knowledge database, and based on it, DIKE built a term extraction model based on Label Specific Attention Network (LSAN)~\cite{xiao2019label} to facilitate the identification of license terms. 
        
        To further enhance the accuracy of license term identification, based on the existing term definitions and classifications from {SPDX}~\cite{SPDX}, {LiDetector}~\cite{lidetector}  
        leverages a probabilistic model and Probabilistic Context-Free Grammar (PCFG) to analyze and categorize customized license content into 23 specific terms. As a subsequent work, {LiResolver}~\cite{LiResolver} incorporates a fine-grained extraction of license entities, relations, and attitudes to achieve detailed fixing of license incompatibilities. Moreover, to automate the extraction of license terms for better readability, they further proposed a more generalized tool {LiSum}~\cite{lisum}, which leverages the correlation between tasks of license text summarization and license term classification to incorporate more information and makes the summarization of licenses more accurate. 

        As another common practice of modifying existing licenses for compatibility issues, Vendome et al.~\cite{machineLearningExcepions} focused on the identification of license exception statements. By studying the prevalent license exceptions in practice, they trained machine-learning-based solutions to automatically identify license exceptions, which make up the empty usage and adoption of OSS license exceptions in this area.


        \noindent \textbf{Industry Perspective:}
        {Most existing SCA tools heavily rely on their pre-built metadata databases for license identification. Specifically, they usually identify potential license risks by identifying the third-party libraries that are integrated first, then query their corresponding licenses by referring to the metadata databases. In this case, it is easy to overlook license-related information that is not maintained in their databases, such as licenses mentioned in code comments~\cite{lidetector}. Among the SCA tools we surveyed in~\Cref{subsec:sca}, only Black Duck~\cite{SCA_BlackDuck7} claims to supplement its analysis by scanning and identifying licenses embedded in software artifacts.}


        \begin{tcolorbox}[colframe=black, colback=white, sharp corners, boxrule=0.5pt, breakable,top=0.2pt, bottom=0.2pt, before skip=3pt, after skip=3pt,left=3pt, right=3pt]
        \textbf{Remark 1:} 
        {There is a huge gap between academic research and existing practices in industrial tools on license identification. Unlike mainstream SCA tools that rely on third-party library metadata for license identification, academic researchers have proposed many works and tools to systematically identify licenses of third-party artifacts, for not only known TPLs but also source code and binaries. Moreover, academic researchers have also noticed the proliferation of customized licenses and tried to propose solutions, and there is an urgent need for industrial tools to incorporate these solutions for better license identification in the future.}         
        \end{tcolorbox}


    \subsection{License Risk {Assessment}}
        Typically there are two major legal issues related to software licenses, 1) software license compatibility, which refers to the ability of different software licenses to coexist in a single project, especially in the scenarios of integrating software artifacts, source code, binary files with different licenses from third-party; 2) software license compliance, which involves adhering to the terms and conditions of software licenses, and this is essential to avoid legal consequences and maintain good standing in the software community.

        \subsubsection{\textbf{License Compatibility}}

        With the widespread practice of software reuse in contemporary software development, it is common for software components from third parties, such as libraries, binaries, and code snippets, to be integrated into the same project to reduce efforts. However, this integration leads to an excessive increase in the risks of license incompatibility issues. To this end, many researchers have investigated the detection of incompatibility issues by various approaches.~\Cref{tab:compatibilityTab} shows the timeline of these research works.
\begin{table}[]
\centering
\caption{A Summary of License Compatibility Detection Research}
\label{tab:compatibilityTab}
\begin{tabular}{clc}
\toprule
\textbf{Method}                                                                    & \textbf{Tool or First Author}                                          & \textbf{Year} \\ \midrule

\multirow{9}{*}{\makecell[c]{\textbf{License Compatibility}\\ \textbf{Analysis Frameworks}}} 

& Nordquist et al.~\cite{firstcompatibility} &      2003 \\
    
    & LChecker~\cite{LChecker}       & 2010 \\
    & Carneades~\cite{Carneades}              & 2011 \\
    
    & Kenen~\cite{Kenen}                      & 2012 \\ 
    & CBDG~\cite{CBDG}                        & 2014 \\ 
    & Ilo et al.~\cite{heterogeneous2015}     & 2015\\
    & Riehle et al.~\cite{riehle2019open} & 2019\\
    & Harutyunyan et al.~\cite{harutyunyan2023open} & 2023\\
    & Sun et al.~\cite{sun2024knowledge} & 2024\\
    \midrule
\multirow{11}{*}{\makecell[c]{\textbf{Incompatible License Pair} \\ \textbf{Based Detection}}}
& German et al.~\cite{codeSiblings}       & 2009 \\
    & German et al.~\cite{auditingLicesning} & 2010 \\
    & Mathur et al.~\cite{GoogleViolation}    & 2012 \\
    & Mlouki et al.~\cite{androidViolation}  & 2016 \\
    & SPDX-VT~\cite{SPDXcompatibility}     & 2017 \\
    & Sorrel~\cite{Sorrel}                 & 2021 \\ 
    & Qiu et al.~\cite{qiu2021empirical}   & 2021 \\
    & Moraes et al.~\cite{multi-licensing_JavaScript_ecosystem}   & 2021 \\
    & Higashi et al.~\cite{dockerCompatibility}     & 2022 \\
    & Makari et al.~\cite{makari2022prevalence} & 2022 \\
    & Antelmi et al.~\cite{antelmi2024analyzing}                 & 2024 \\
    \midrule
\multirow{8}{*}{\makecell[c]{\textbf{Term Conflict} \\ \textbf{Based Detection}}}

    & Alspaugh et al.~\cite{heterogeneous2012Alspaugh}    & 2012  \\
    & LicenseRec~\cite{LicenseRec}             & 2023  \\
    & SILENCE~\cite{pypi}                      & 2023  \\
    & LiDetector~\cite{lidetector}             & 2023  \\ 
    & LiResolver~\cite{LiResolver}             & 2023  \\ 
    & DIKE~\cite{DIKE}                         & 2023  \\
    & Liu et al.~\cite{catchTheButterfly}      & 2024  \\
    & Wu et al.~\cite{baolingfenglicense} & 2024 \\
    \bottomrule
\end{tabular}
\end{table}

        \noindent $\bullet$ \textbf{{License Compatibility Analysis Frameworks}}: 
        {The issue of license compatibility has already been recognized and analyzed by researchers since very early time. Although there was initially a lack of consensus and sufficient data regarding license incompatibility, researchers have proposed analytical frameworks to capture the scenarios of license adoption and possible conflicts.
        Nordquist et al.~\cite{firstcompatibility} are the very first to propose tools that intercept the build process and identify essential dependencies to examine to identify the applied software licenses, therefore, flagging missing licenses or potential conflicts. 
        Van Der Burg et al.~\cite{CBDG} then refined the dependencies by tracing system calls during builds, they constructed a Concrete Build Dependency Graph (CBDG) to identify sources that are truly included in deliverables and examine license compatibility issues within them.}
        {
        Moreover, considering that developers gradually learn and reuse existing open-source software in their own projects,  
        Zhang et al.~\cite{LChecker} proposed LChecker, which utilizes Google Code Search service to check whether a local file exists in an OSS project and whether the licenses are compatible.
        German et al.~\cite{Kenen} systematically discussed the challenges in license compliance, and based on that, introduced {Kenen},  a semiautomatic process for license incompatibility analysis for Java projects.}
        
        {Some researchers also contributed to the formalization and modeling of license compatibility issues.
        Gordon~\cite{Carneades} introduced a prototype legal application for analyzing OSS license compatibility issues based on {Carneades},
        which modeled and formalized legal rules and reason possible license incompatibility issues by OWL.}
        {Ilo et al.~\cite{heterogeneous2015} also proposed SWREL, an ontology-based semantic modeling approach to combine and consequently query information about software interrelationships across different ecosystems, based on these, they managed to detect potential license violations with user dependencies. 
        With the evolution of software supply chains, researchers have introduced new governance frameworks~\cite{riehle2019open}, prompting organizations and projects to adopt standardized license management processes~\cite{harutyunyan2023open, sun2024knowledge}. These frameworks primarily focus on establishing comprehensive SBOMs and strengthening audits of third-party components, thereby mitigating the risk of internal license conflicts arising from the integration of upstream software components.}



        {
        }

        \noindent $\bullet$ \textbf{Incompatible License Pair Based Detection}:
        {With the growing awareness and adoption of OSS licenses, both academic and industrial researchers have contributed a lot to building the consensus understanding of incompatibility among mainstream licenses. Based on well-established incompatible license pairs, }
        German et al.~\cite{codeSiblings} investigated the potential license violations during the frequent code exchange between Linux and two BSD systems with Unix kernels (i.e., FreeBSD and OpenBSD). 
        {Similarly, Mathur et al.~\cite{GoogleViolation} evaluated the code migration on Google Code and examined their potential license compatibility issues. 
        }
        {Particularly, due to the viral nature of copyleft, GPL family licenses are mostly concerned by researchers and industrial SCA products~\cite{SCA_BlackDuck9,SCA_Mendio3,SCA_GitLab2,SCA_Aqua2}.}
        {German et al.~\cite{auditingLicesning} analyzed inconsistencies between the declared licenses of binary files and source code in the Fedora 12 operating system, to identify GPL-related compatibility issues.}
        {Mlouki et al.~\cite{androidViolation} investigated the licenses that are mostly used in mobile apps, and how these licenses, as well as potential license violations, especially GPL-related incompatibility, evolve over time.} 
        %
        %
        {Higashi et al.~\cite{dockerCompatibility} analyzed Docker images from GitHub and found that GPL-incompatible software packages are fairly common, with a significant portion of images exhibiting these issues.}
        
        {As the proliferation of OSS licenses, apart from GPL licenses, researchers also gradually concentrated on compatibility issues introduced by other new-emerging licenses. 
        Kapitsaki et al.~\cite{SPDXcompatibility} proposed the compatibility map, which largely expanded the scope of compatibility detection. They developed a relation graph for over 20 prevalent licenses, detailing features like link types, license upgrade options, and one-way compatibility. 
        Moraes et al.~\cite{multi-licensing_JavaScript_ecosystem} also utilized this graph to conduct an empirical analysis in JavaScript projects, they found that 24\% multi-licensed projects introduced at least one license incompatibility issue.} 
        Following this principle, 
        {SORREL}~\cite{Sorrel} is proposed to retrieve licenses from metadata and assess incompatibilities for Java projects by referencing predefined compatibility lists. 
        Qiu et al.~\cite{qiu2021empirical} also conducted an empirical study on the incompatibility issues hidden in the dependency relations among NPM packages based on a license compatibility graph of 17 popular licenses.
        Makari et al.~\cite{makari2022prevalence} integrated license incompatibility information from multiple sources and constructed a matrix containing 1,476 license compatibility pairs (i.e., among approximately 40 licenses). 
        {Based on this, they conducted empirical studies on compatibility in the NPM and RubyGems ecosystem. Compared to Qiu et al.~\cite{qiu2021empirical}, they identified much more incompatibility issues due to higher license coverage.}
        {Similarly, Antelmi et al.~\cite{antelmi2024analyzing} introduced the Software Heritage Analytics (SWHA) framework to look into the license compatibility issues, they also followed the existing OSADL license checklist~\cite{OSADL} to facilitate comprehensive analysis of license incompatibility among GitHub repositories.}

        \noindent $\bullet$ \textbf{Term Conflict Based Detection}: 
        Due to the proliferation of customized licenses, license matrix-based detection is becoming less adequate for practical requirements. {For instance, SPDX License List~\cite{SPDX} indexed the popular licenses that are widely adopted in practice, which is far more than those maintained by any license compatibility knowledge.} In this case, term-conflict-based detection is becoming the mainstream technique in license incompatibility analysis.

        Some researchers focused on establishing a more detailed and comprehensive knowledge base of popular license compatibility based on term conflict. Xu et al.~\cite{LicenseRec} considered 19 license terms to build a compatibility matrix~\cite{compatibilityMatrix}, which contains one-way compatibility information for 63 types of licenses certified by {FSF} or {OSI} in their license recommendation tool {LicenseRec}. In their subsequent tool, {SILENCE}~\cite{pypi}, they implemented a breadth-first algorithm imitating the search behavior of PyPI dependency resolution to get dependencies in projects and combined the compatibility matrix to detect license compatibility in the PyPI ecosystem. 
        
        {
        Some researchers also directly examined license conflict based on detailed license terms instead of whole licenses.
        Alspaugh et al.~\cite{alspaugh2009analyzing} first identified key properties of OSS licenses, presented a license analysis scheme, and proposed the semantic parameterization analysis of open-source licenses to fulfill intellectual property requirements management for heterogeneously-licensed systems~\cite{HeterogeneouslyLicensedSystems2009Alspaugh}. Moreover, to better explain the license conflicts for software architects, they further proposed a tool to simultaneously provide explanations, as well as the trade-offs for alternative licenses, in heterogeneously-licensed systems~\cite{heterogeneous2012Alspaugh}. }
        %
        
        {Xu et al.~\cite{lidetector} deepened the idea of license modeling and introduced an automated term-conflict-based detection tool, LiDetector. {Based on this, they further defined term conflicts in detail by considering different licensing scopes, i.e., parent-child license incompatibility~\cite{LiResolver}.}
        {Wu et al.~\cite{baolingfenglicense} utilized {LiDetector} and conducted an empirical analysis of license usage across five major package management platforms (Maven, NPM, PyPI, RubyGems, and Cargo). The fine-grained, term-based detection method revealed that approximately 8\% of cases exhibited license incompatibilities, primarily due to the violation of sublicensing~\cite{Sublicensing} between GPL and permissive licenses.}
        Cui et al.~\cite{DIKE} further extended the incompatibility detection to a wider range of 3,256 licenses, they developed {License Analyzer} to automatically retrieve license terms and reasoning potential license incompatibility by \textit{License-Terms-Responsibility} relationships.
        Liu et al.~\cite{catchTheButterfly} enhanced term-conflict-based detection from the perspective of term framework, they conducted a term-level analysis of term labeling on 453 {SPDX} licenses. They summarized the inconsistencies in license terms from Choosealicense~\cite{choosealicense}, TLDRLegal~\cite{tldrlegal} and OpenEuler~\cite{LicenseShowRoom}, and marked these {SPDX} licenses with refined terms and extended conflict relationships.} 
        

        
        \noindent \textbf{Industry Perspective:}
        {Academic studies on license incompatibility analysis focus more on the complex and context-aware models that aim to provide detailed analysis of license interactions to comply with legal and ethical requirements, while industrial SCA tools typically focus on detecting basic license conflicts or flagging high-risk licenses, based on existing known knowledge. Especially, leading SCA tools~\cite{SCA_BlackDuck9,SCA_Mendio3,SCA_GitLab2,SCA_Aqua2} focus more on license conflicts that could influence commercial usages by industry, such as the well-known viral copyleft licenses~\cite{virallicense}. To achieve better license incompatibility detection, Mend.io~\cite{SCA_Mendio3} has adopted license compatibility graphs and tables as references to identify a broader range of license incompatibilities, while these predefined compatibility resources are relatively limited.}
        


        \begin{tcolorbox}[colframe=black, colback=white, sharp corners, boxrule=0.5pt, breakable,top=0.2pt, bottom=0.2pt, before skip=3pt, after skip=3pt,left=3pt, right=3pt]
        \textbf{Remark 2:}
        {There is a huge gap in license incompatibility analysis between academia and industry, academic researchers have conducted numerous works to identify any possible license incompatibility (even if license terms are not detected in a standardized license), while industrial SCA tools mainly focus on identifying certain types of license incompatibility that could really lead to commercial losses. This could be due to the unawareness of rights and obligations in open-source communities.
        We believe there is still a long way for different stakeholders to reach consensus on the severity of license incompatibility.}
        \end{tcolorbox}

        \subsubsection{\textbf{License Compliance}}

\begin{table}[]
\centering
\caption{A Summary of License Compliance Detection Research}
\label{tab:violationTab}
\begin{tabular}{cp{4cm}r}
\toprule
\textbf{Categories} & \textbf{Tool or First Author} & \textbf{Year} \\
\midrule
\multirow{12}{*}{{\textbf{\textbf{Source Code}}}  }

  & Monden et al.~\cite{cloneMetrics}   & 2010   \\
  &Boughanmi~\cite{heterogeneous2010} & 2010\\

  & An et al.~\cite{ToxicAndroid}   & 2017   \\
  &Lotter et al.~\cite{stackoverflowJava} &2018 \\
  & Baltes et al.~\cite{ToxicGitHub}   & 2019   \\  
  
  & Ragkhit. et al.~\cite{ToxicOnStackOverflow}   & 2019   \\ 
  & Golubev et al.~\cite{GitHubViolation}   & 2020   \\
  & Serafini et al.~\cite{serafini2022efficient} & 2022\\
  & Ciniselli et al.~\cite{deeplearningLicense} & 2022\\
  & CODEIPPROMPT~\cite{CODEIPPROMPT} & 2023\\
  &Kapitsaki et al.~\cite{kapitsaki2024generative} &2024\\
   & ModelGo~\cite{duan2024modelgo} &2024\\
  
\midrule
\multirow{5}{*}{{\textbf{\textbf{Binary}}}}
  & Molina et al.~\cite{molina2007fast}  & 2007\\
  & BAT~\cite{BAT}   & 2011   \\
  & OSSPolice~\cite{OSSPolice}   & 2017   \\
  & Feng et al.~\cite{BinaryViolation}   & 2019   \\
  & OSLDetector~\cite{OSLDetector} & 2020\\

\midrule
\multirow{8}{*}{{\textbf{\textbf{License Tampering}}}} 
  & Di Penta et al.~\cite{di2009source} & 2009\\
  & Di Penta et al.~\cite{di2010exploratory} & 2010\\
  & Wu et al.~\cite{inconsistenciesInLarge-scaleProjects}  & 2015    \\
  & Vendome et al.~\cite{vendome2015licensechange} &2015\\
  & Mlouki et al.~\cite{androidViolation}   & 2016   \\
  & Wu et al.~\cite{inconsistencyInLargeCollections}   & 2017    \\
  & Vendome et al.~\cite{vendome2017licensechange} &2017\\
  &Reid et al.~\cite{reid2023applying} &2023\\

\bottomrule
\end{tabular}

\end{table}

        Besides license incompatibility among licenses, detecting the overall compliance of all licenses before officially releasing is also vital for software providers. 
        \Cref{tab:violationTab} presents the existing research on software license non-compliance detection.
        
        \noindent $\bullet$ \textbf{License Non-compliance Detection on Source Code}: 
        The most important concept in the context of license non-compliance lies in the definition of \textit{derivative works}. Specifically, to what extent source code reuse can be identified as \textit{derivative works}, {there is still no consensus so far.
        To bridge this gap, most research works rely on code clone detection techniques to distinguish the potential reuse of protected intellectual property of source code.}
        
        {Monden et al.~\cite{cloneMetrics} were the first to convene expert programmers to empirically analyze source code reuse under the GPL license, they introduced metrics such as the Maximum Length of a Clone (MLC) and the Local Similarity Lower Bound (LSim), to facilitate the identification of license violations. 
        Based on this, 
        Golubev et al.~\cite{GitHubViolation} investigated the potential code borrowing among Java projects on GitHub and unveiled that 29.6\% of these Java projects might be involved in potential code borrowing and 9.4\% of them could potentially violate original licenses.
        Serafini et al.~\cite{serafini2022efficient} integrated Software Heritage Project and introduced SWH-scanner to efficiently identify prior publications existing in user projects, to alert where it comes from and whether additional obligations shall be fulfilled before product shipment.}
        
        
        
        {
        Apart from GitHub, developers also refer to online technical forums, such as Stack Overflow (SO), to learn and reuse existing solutions. However, 
        such reuse could also lead to licensing challenges.}
        {When reusing code from Stack Overflow, many developers ignore its \textit{CC BY-SA} requirements~\cite{SOCCBYSA}, and some also fail to comply with the original project's licensing when posting its snippets on the platform.}
        {An et al.~\cite{ToxicAndroid} first conducted a large-scale empirical study on this phenomenon, revealing extensive bi-directional code reuse between the Stack Overflow platform and Android projects. However, nearly all instances of code reuse failed to adhere to the requirements of proper license declarations.}
        {Lotter et al.~\cite{stackoverflowJava} conducted a similar investigation between Java projects and SO, revealing the prevalence of code reuse in these copying flows. They also found that a significant portion of reused code lacked proper attribution, further increasing the risks associated with merging such code.}
        %
        {Baltes et al.~\cite{ToxicGitHub} further conducted a large-scale empirical study and surveys to investigate to what extent developers are aware and follow the attribution requirement by SO, they found that almost one-half admitted copying code from SO without attribution and about two-thirds were not aware of the license of SO code snippets and its implications.}
        
        %
        Moreover, since Stack Overflow also allows users to license their submitted content, Ragkhitwetsagul et al.~\cite{ToxicOnStackOverflow} extended to the compliance of these user-defined licenses and noticed that many source code snippets that are under copyleft licenses, are submitted as Stack Overflow answers while being relicensed with permissive licenses. In this case, such non-compliant relicense could also risk downstream users and lead to potential legitimate issues.

        {
        As open-source projects increasingly involve diverse forms of code reuse and integration, determining whether a codebase constitutes \textit{derivative work} may not be solely addressed through clone detection methods. Boughanmi~\cite{heterogeneous2010} proposed a compliance analysis framework that summarizes the interconnection mechanisms between components, such as linking, fork, subclass, IPC, or plugin, while constructing the dependency graph of the software system. These interconnection types may influence the compliance assessment for derivative projects. 
        Duan et al.~\cite{duan2024modelgo} focused on the scenario of deep learning projects, in which it involves interactions among diverse types of licenses
        and licensed materials for compliance checks. To this end, they developed ModelGo, a practical tool to handle complex interactions and audit
        potential legal risks in machine learning projects to enhance compliance and fairness.}
        
        
        {
        With the gradual engagement of AI-based techniques in modern software development, it is unavoidable that LLMs that are trained from open-source data, can easily generate code that is originally from open-source projects to users, which makes it even blurred to distinguish the \textit{derivative works} in the context of license compliance. To this end, some researchers have tried to unveil the potential risks behind generative software engineering.  
        Ciniselli et al.~\cite{deeplearningLicense} first proposed the concern that it is unclear whether the code generated by DL models trained on open-source code should be considered as new or as derivative work, with possible implications on license infringements. They investigated the extent to which DL models tend to clone code from their training set when recommending code completions. They found that Type-1 clones of training instances account for approximately 10\% of short predictions, dropping to 0.1\% for longer ones.
        To specifically evaluate the potential IP violation risk, Yu et al.~\cite{CODEIPPROMPT} proposed CodeIPPrompt, which automatically evaluates the extent to which code language models may reproduce licensed programs. They found that most IP violation issues in generated code are because of the implicit inclusion of restrictive code in permissive and public-domain code repositories due to inconsistent licensing practices. 
        To avoid potential legitimate risks, Kapitsaki et al.~\cite{kapitsaki2024generative} systematically discussed the responsible model, as well as possible legitimate issues that could be introduced by potential code reuse, along the workflow of LLM-based code generation tasks.}
        \noindent $\bullet$ \textbf{License Non-compliance Detection on Binary}: 
        Binary is another typical type of deliverables of software artifacts. Compared to detecting source code reuse, it is usually not that straightforward to detect third-party reuse in binaries, and such reuse is usually not accompanied by explicit license declarations, making it easy to overlook and challenging to detect potential non-compliance.

        {Molina et al.~\cite{molina2007fast} were the first to propose using binary code cloning techniques for detecting unauthorized use of open-source projects in pirated programs.
        They transformed binary programs into Static Single Assignment (SSA) and then extracted expression trees as fingerprints for similarity matching.
        Besides reverse engineering, some researchers also map binaries back to source code to associate the binaries to their licenses for compliance assessment.}  
        {Hemel et al.~\cite{BAT} proposed {BAT} to inspect string literals that are usually invariant during compilation, as cursors, to identify potential code clones from third-party binaries or source code.}
        To further enhance the accuracy and efficiency of similarity analysis, OSSPolice~\cite{OSSPolice} incorporated the inherent characteristic features of targeted binaries, such as string literals, constants, normalized classes, and exported functions, in the similarity comparison. By comparing to a well-constructed dataset of 132K OSS sources, OSSPolice revealed that over 40K apps potentially violate GPL/AGPL licensing terms. 
        Feng et al.~\cite{BinaryViolation} extended the similarity comparison by comprehensively including not only code features that are invariant during compilation and could vary between projects, but also file attributes of executable files, such as file names, internal names, company names, and legal copyright, from \textit{VERSIONINFO} resource of PE files.
        {However, these more general features could easily introduce false positives in code clone detection.
        To deal with this, Zhang et al.~\cite{OSLDetector} proposed OSLDetector, based on internal cloning forest, which improves the efficiency of feature duplication between libraries, resulting in the optimal selection of clone detection candidate as the final result.}
        
        



        \noindent $\bullet$ \textbf{License Tampering:}
        The license tampering in OSS presents a significant challenge in identifying potential license risks. The widespread practice of reusing existing open-source code through direct copy-pasting often leads to a single code snippet being traced back to multiple open-source repositories when determining third-party reuse. However, many instances of such copying neglect the critical necessity of verifying the compliance requirements of the original licenses (i.e., missing necessary copyright and license notice, or directly relicensing). 

        To deal with this challenge, researchers are gradually concerned with the potential threats of license provenance changes. 
        {Di Penta et al.~\cite{di2009source, di2010exploratory} proposed an automated approach for license tracking. They extracted change information from version control systems as input, identified license content within these changes, and conducted empirical studies to summarize the evolving status of contributors, types of licenses, and copyright years across multiple open-source repositories.
        By examining such metadata information, researchers also unveil the original licenses and changes of libraries and code snippets in different ecosystems. 
        Researchers have also extended such license consistency analysis to various scenarios, 
        including the Debian project~\cite{inconsistenciesInLarge-scaleProjects}, Java repositories on GitHub~\cite{inconsistencyInLargeCollections}, and the Android ecosystem~\cite{androidViolation}.
        License tampering also happens in version releases, {Vendome et al.~\cite{vendome2015licensechange, vendome2017licensechange} analyzed license changes in the version evolution of GitHub projects and found that many changes lack corresponding commit records or discussions, resulting in poor traceability.}
        Reid et al.~\cite{reid2023applying} further confirmed the prevalence of license omission significantly introduced license violations at a large scale by tracking the license altering history logged in World of Code, suggesting the necessity of regulations of license traceability for open-source reuse.}
        

        
        \noindent \textbf{Industry Perspective:}
        {To identify potential license compliance risks in projects, the industry similarly emphasizes code tracing analysis. However, this domain rarely focuses on detecting the fulfillment of obligations specified by license terms. Instead, they focus more on identifying whether specific high-risk license types of interest to developers appear in the project. Additionally, with the rise of generative code, some leading SCA tools have realized potential compliance risks associated with generative code~\cite{SCA_BlackDuck4,SCA_Checkmarx3}. Nevertheless, current detection methods remain limited, relying on component analysis of generated code to monitor compliance risks.}

        \begin{tcolorbox}[colframe=black, colback=white, sharp corners, boxrule=0.5pt, breakable,top=0.2pt, bottom=0.2pt, before skip=3pt, after skip=3pt,left=3pt, right=3pt]
        \textbf{Remark 3:} 
        {There is a gap between academia and industry in the assessment of license compliance. Unlike mainstream SCA tools, which focus solely on identifying high-risk license types within a project, academic research emphasizes whether specific license obligations are adhered to across different code usage scenarios. In the context of generative code, academic works have extended beyond traditional code tracing to include the analysis of prompts and the code from training datasets. These phenomena highlight the industry's cautious reliance on manual analysis for compliance detection. However, given the increasing complexity of code application scenarios and license terms, more detailed compliance analysis approaches need to be integrated for industry advancements.}
        \end{tcolorbox}
        

    \subsection{{License Risk Mitigation}}
        {
        In this section, we discuss the countermeasures proposed by academic research, as listed in~\Cref{tab:countermeasureTab}.}

        \begin{table}[]
\centering
\caption{{A Summary of License Risk Mitigation Research}}
\label{tab:countermeasureTab}

\resizebox{0.49\textwidth}{!}{

\begin{tabular}{c|clr}

\Xhline{0.8px}
\textbf{Categories} & \textbf{Objective} & \textbf{Tool or First Author} & \textbf{Year}\\
\hline
\multirow{14}{*}{\textbf{Risk Remediation}} &\multirow{14}{*}{\makecell[c]{-}} 

& German et al.~\cite{intetrationPatterns} & 2009\\
&  & Hammouda et al.~\cite{hammouda2010open}  & 2010\\
&  & Scacchi et al.~\cite{scacchi2012understanding}  & 2012\\
&  & Viseur~\cite{selectionofCloudComputing} & 2016\\
&&Vendome et al.~\cite{vendome2016assisting}& 2016\\
&&Mlouki et al.~\cite{androidViolation}&2016\\
&&findOSSLicense~\cite{FindossLicense2}&2019\\
&&ALP~\cite{ALP}&2019\\
&&Sorrel~\cite{Sorrel}&2021\\
&&LicenseRec~\cite{LicenseRec}&2023\\
&&DIKE~\cite{DIKE}&2023\\
&&LiResolver~\cite{LiResolver}&2023\\
&&SILENCE~\cite{pypi}&2023\\

&& LiVo~\cite{huang2024your} & 2024\\

\hline

\multirow{15}{*}{\textbf{\makecell[c]{Recommendation\\and Selection}}}

 & \multirow{11}{*}{\makecell[c]{Concerns on OSS\\License Selection}} 
&Colazo et al.~\cite{colazo2005copyleft}&2005\\
 &&Stewart et al.~\cite{stewart2005preliminary}&2005\\
&&Colazo et al.~\cite{colazo2009impact}&2009\\
&&Lindman et al.~\cite{lindman2011matching}&2011\\
&&Kashima et al.~\cite{kashima2011investigation}&2011\\
&&Ghapanchi et al.~\cite{ghapanchi2011impact}&2011\\
&&Vendome et al.~\cite{vendome2015licensechange}&2015\\
&&Vendome et al.~\cite{developerRef2}&2015\\
&&Vendome et al.~\cite{vendome2017licensechange}&2017\\
&&Medappa et al.~\cite{medappa2017license}&2017\\
&&Gamalielsson et al.~\cite{gamalielsson2017licensing}&2017\\

   \Xcline{2-4}{0.5px}

   & \multirow{4}{*}{\makecell[c]{Approaches to Narrow\\down License Selection}} 
&ALP~\cite{ALP}&2019\\
&&findOSSLicense~\cite{FindossLicense2}&2019\\

&&LicenseRec~\cite{LicenseRec}&2023\\
&&Zhang et al.~\cite{recommendationPower}&2024\\

\Xhline{0.8px}
\end{tabular}

}

\end{table} 

        \subsubsection{{\textbf{License Risk Remediation}}}



        To systematically eliminate identified licensing risks, researchers have conducted empirical studies on solutions that have emerged in practice. 
        After investigating
        solutions to license incompatibilities in 124 OSS packages, German et al.~\cite{intetrationPatterns} identified twelve categories of remedial strategies, which are applicable to both licensors and licensees, such as loosening license restriction for licensors, modifying code, adjusting licenses, and negotiation with licensors to resolve potential license conflicts.
        {Hammouda et al.~\cite{hammouda2010open} proposed the concept of open-source legality patterns in detail, and introduced a number of open-source legality patterns identified in real systems and expert interviews, such as decoupling component interactions, isolating critical business logic, and adjusting licensing schemes, to avoid potential license incompatibility in real-world software architectural design.
        Scacchi et al.~\cite{scacchi2012understanding} also argued that the coupling relationships between components in open architecture systems can be leveraged to mitigate the risks of license incompatibilities, such as component replacement, architectural refactoring (e.g., switching from static linking to dynamic linking), and license updates.}
        Mlouki et al.~\cite{androidViolation} surveyed solutions for resolving conflict between project-declared licenses and code internal licenses across 859 Android applications. They find that developers typically address these issues by changing the licenses or removing the contentious files.

        {
        Some researchers also turned these possible operations into automated solutions, to remediate identified legitimate issues as much as they can.
        This line of research advanced alongside compatibility detection, evolving from conceptual proposals~\cite{selectionofCloudComputing, vendome2016assisting, ALP} to incompatible license pair-based detection~\cite{FindossLicense2, Sorrel}, and ultimately to the evaluation of specific license terms~\cite{LicenseRec, DIKE}.}
        However, this automated approach is limited as there is not always a possible compatible alternative to perfectly resolve the conflicts.
        In this case, {LiResolver}~\cite{LiResolver} was proposed to address this by customizing license terms. LiResolver also seeks licenses that eliminate term conflicts as replacements first. When such a license is unavailable, {LiResolver is also the only tool that} customizes the terms of an existing license, such as converting the term attitudes of the constraint requirements to the most stringent option or generating exceptions to the license, to resolve conflicts. 
        This paper also mentioned that if the conflicting components are developed by the same developer, replacing the license directly could be considered.
        Xu et al.~\cite{pypi} further extended license risk remediation by library migration techniques. They first conducted an empirical study on common remediation methods for incompatibilities in the PyPI ecosystem, identifying five popular strategies: \textit{Migration}, \textit{Removal}, \textit{Pin Version}, \textit{Negotiation}, and \textit{Change Own License}.
        They collected data on migrated packages from their previous work~\cite{migrationPackage} and license information across different package versions, then used an SMT solver to determine if conflict resolution could be achieved by replacing conflicting components.
        {Huang et al.~\cite{huang2024your} also delved into the issue of missing modification notices which are one of the most common causes of license non-compliance. They proposed LiVo to detect gaps in required notices within GitHub commit logs and supplement missing details to ensure compliance. {LiVo is also the only work to date that resolves missing modification declarations.}}

        \noindent \textbf{Industry Perspective:}
        {The mitigation of identified license risks is relatively less concerning in industrial SCA tools. Typically, they only either leverage metadata to automatically generate attribution lists for third-party dependencies to address missing attribution issues~\cite{SCA_Sonatype2,SCA_BlackDuck5,SCA_Mendio4,SCA_Revenera3}, or simply provide general suggestion for manual remediation in generated risk reports~\cite{SCA_BlackDuck4,SCA_Revenera4,SCA_Checkmarx4, SCA_Palo2,SCA_Sonatype4,SCA_Mendio2}, which heavily rely on manual review and operations. However, it is non-trivial for developers to understand the rights and obligations behind legal terms in generated reports, which could largely hinder the mitigation of license-related risks.}


        \begin{tcolorbox}[colframe=black, colback=white, sharp corners, boxrule=0.5pt, breakable,top=0.2pt, bottom=0.2pt, before skip=3pt, after skip=3pt,left=3pt, right=3pt]
        \textbf{Remark 4:}
        {Compared to academic research that mostly focuses on the automation of license risk mitigation, industrial tools are much more cautious and careful when introducing automated solutions. However, simply leaving users alone with detailed reports with general suggestions also does not really solve the challenge of developers who are not familiar with legitimate issues. To this end, the issue of how to reach an agreement between automation and acceptable manual validation should be reconsidered by both academic researchers and industrial vendors.}        
        \end{tcolorbox}



    \subsubsection{{\textbf{License Recommendation and Selection}}}
    {Apart from remediating identified legitimate issues, how to select and assign proper licenses to user projects is another widely concerned scenario by developers. To this end, researchers have also delved into the recommendation and selection of licenses to developers.}

        
        \noindent $\bullet$ {\textbf{Concerns on OSS License Selection:}}
        {The selection of an appropriate open-source license plays a key role in promoting sustainable software development.~\cite{colazo2009impact}.
        Empirical studies investigating the impact of license types on open-source projects have analyzed diverse scenarios, encompassing widely used projects on GitHub~\cite{vendome2015licensechange, vendome2017licensechange, medappa2017license}, large-scale repositories from Debian and SourceForge~\cite{colazo2005copyleft, kashima2011investigation}, projects supported by companies~\cite{stewart2005preliminary, ghapanchi2011impact, medappa2017license}, and the top-ranked OSS projects on OpenHub~\cite{gamalielsson2017licensing}. These studies collectively reveal that permissive licenses are particularly favored due to their ability to attract commercial sponsorship, increase developer engagement, and facilitate code reuse across diverse contexts. In contrast, copyleft licenses, while potentially limiting user interest and project vitality, provide distinct advantages such as better safeguarding openness, fostering stronger alignment between developers and their communities, and attracting intrinsically motivated contributors.
        }
    

        {Apart from examining how license strictness affects project success, other research has explored additional factors that shape license decision-making. Vendome et al.~\cite{developerRef2} analyzed 16,221 Java projects and surveyed 138 developers, identifying critical drivers such as compatibility risks, commercial reuse, and user feedback. 
        Lindman et al.~\cite{lindman2011matching} analyzed related factors from a commercial perspective and, through case studies and interviews, identified key drivers such as externalities (i.e., compatibility with existing systems), developer motivation, community leadership, and company size.
        }
        


        \noindent $\bullet$ {\textbf{Approaches to Narrow down License Selection:}}
        {Some researchers also focus on developing assistant tools to help users narrow down and select appropriate licenses for their own projects.}
        LicenseRec~\cite{LicenseRec} asked users seven questions to obtain their requirements on licenses, then filtered out suitable licenses by matching terms recommended to users by referring to their knowledge base~\cite{compatibilityMatrix}.
        
        Apart from user needs, some researchers also investigated the practice of considering similar projects as a basis for license recommendation, in various scenarios.
        Zhang et al.~\cite{recommendationPower} focused on license selection for software in the electric power industry. Considering that it could be difficult for non-developers to understand the complex concepts and threats of software licenses, they rely on project information (e.g., the project description, readme text, chosen topics, programming language, project size, and the year) to cluster existing power-related GitHub repositories, and recommend software licenses to these users to simplify their license selection.
        
        ALP~\cite{ALP} incorporates the project's historical versions and change information into license selection considerations. It considers various project features to recommend licenses, including license-related content in code-inline text and changed text, the impact of project historical versions on license selection, the content of project document texts, and the license differences in co-changed files. The study uses the Conditional Random Field (CRF)~\cite{crf} algorithm to predict licenses based on different features respectively. If there are discrepancies in the predictions, a ranking-learning algorithm provided by libSVM~\cite{libsvm} is used to rank the features contributing to the discrepancies, with the highest-ranked instance being used as the basis for the recommended license.

        {FindOSSLicense}~\cite{FindossLicense2} refined its license selection approach by integrating both user and project properties into its recommendations. By evaluating 33 well-known licenses, they first identified 38 distinct license terms, and further gathered user requirements through questionnaires, such as user demand on license restrictions.
        These insights helped filter license terms to generate preliminary recommendations. After that, the tool factored in the programming language and application type of user projects and recommended licenses that align with these properties. Lastly, findOSSLicense utilized a collaborative filtering algorithm that leveraged these properties and user preferences for licenses to make final recommendations.

        \textbf{Industry Perspective:}
        {Unlike academic research that has delved into the selection of licenses from various aspects concerned by different stakeholders, industrial SCA tools seldom recommend licenses to user projects. After going through all leading SCA tools in~\Cref{tab:SCAtoolFunctions}, only JFrog~\cite{SCA_JFrog3} and GitHub~\cite{SCA_GitHub3} specifically give recommendations to repositories that have no license assigned, while their recommendations are also straightforward without additional guidance to assist with the license selection process.} 
        

        \begin{tcolorbox}[colframe=black, colback=white, sharp corners, boxrule=0.5pt, breakable,top=0.2pt, bottom=0.2pt, before skip=3pt, after skip=3pt,left=3pt, right=3pt]
        \textbf{Remark 5:}       
        {There is an obvious delay in the adaption of license recommendation approaches to the industry. Academic researchers have investigated a lot in the factors that could influence the license selection, while only a few of them are incorporated in existing SCA tools. This also unveils that existing SCA tools are less interested in solving license missing problems in user projects, which should be further enhanced to complete the lifecycle of license management in SCA tools.}
        \end{tcolorbox}

\section{Discussion}\label{sec:discussion}
    {In this section, we
    discuss the existing challenges in the procedures, and outline potential opportunities for future development in this area. Additionally, we also provide practical recommendations for practitioners to enhance their license management processes.}

    \subsection{Challenges in Current License Risk Management}
        \newcounter{discussionCount} 
\stepcounter{discussionCount}
\noindent \textbf{Challenge \arabic{discussionCount}: {License Proliferation and Lagging Tool Support.}}
{With the increasing specified requirements of project management, existing licenses are gradually customized and proliferated to satisfy maintainers' needs.}
{However, such customizations could introduce rights and obligations that are not commonly defined in existing mainstream software licenses, leading to the increasing difficulty for existing license management tools to identify, parse, and ensure their compatibility and compliance with other licenses.}


{It is evident that maintaining a comprehensive and up-to-date license knowledge base is a critical foundation for effective license management methods and tools. However, license proliferation often leads to fragmented, flawed, or outdated knowledge, diminishing the practical utility of these methods and tools.}
{As pointed out by OSI, the extensive growth of OSS licenses (i.e., license proliferation) not only complicates the selection process for licensors but also leads to incompatibilities and confusion over the terms in multi-license distributions~\cite{osiProliferation}.}
For instance, both rule-based and fuzzy matching methods for license identification necessitate regular updates and meticulous maintenance to remain effective. Their success hinges critically on the completeness and accuracy of the license rule sets employed.

\begin{table}[]
\centering
\caption{{Number of licenses identified by Identification Methods}}
\label{tab:numofIdentification}


\begin{tabular}{l|c|c}
\Xhline{0.8px}
\textbf{Tool or First Author}                                      & \textbf{Database}                   & \textbf{\#License}            \\ \Xhline{0.8px}
Wolter et al.~\cite{inconsistenciesOnGithub} & ScanCode LicenseDB                  & 2000+                         \\ \hline
Nomos~\cite{fossology}                       & \multirow{3}{*}{FOSSology Database} & \multirow{3}{*}{1700+}        \\
Monk~\cite{fossology}                        &                                     &                               \\
Wolter et al.~\cite{inconsistenciesOnGithub} &                                     &                               \\ \hline
Wu et al.~\cite{baolingfenglicense}          & SPDX License List                   & 606                           \\ \hline
Ninka~\cite{ninka}                           & \multirow{3}{*}{Ninka Database}     & \multirow{3}{*}{112}          \\
Higashi et al.~\cite{clusterninka2}          &                                     &                               \\
Higashi et al.~\cite{clusterninka3}          &                                     &                               \\ \hline
ASLA~\cite{ASLA}                             & ASLA Database                       & 37                            \\ \hline
SORREL~\cite{Sorrel}                         & SORREL Database                     & 16                            \\ \hline
Di Penta et al.~\cite{jarArchives}           & \multirow{2}{*}{Google Code Search} & \multirow{2}{*}{\boyuan{Inaccessable}} \\
LChecker~\cite{LChecker}                     &                                     &                               \\ \Xhline{0.8px}
\end{tabular}


\end{table}

{We conducted an in-depth analysis of collected papers to ascertain the scope of license types encompassed by various identification tools, as outlined in these papers and their related tool documentation.}
{As presented in~\Cref{tab:numofIdentification}, existing license identification techniques significantly vary on the scope of software licenses they support. However, the covered licenses are still far from real-world license adoptions. According to Jesus M. et al.~\cite{gonzalez2023software}, 6.9 million unique license declarations are identified from Software Heritage. Although there could be many variations and exceptions of mainstream licenses~\cite{lidetector, LiResolver}, we believe the scope of important licenses is still far larger than current support.}
{Moreover, the effective license risk analysis also heavily relies on the reliable identification of granted software licenses. For instance, many existing tools rely on well-built compatibility databases to detect license violations, while the most advanced compatibility matrices only cover over 100 licenses~\cite{compatibilityMatrix, compatibilityMatrix2}, which is far from comprehensive coverage for the proliferated licenses and require massive efforts to make up.
To this end, with the increasing growth of customized licenses, it could be even more challenging to sustain these knowledge-based solutions for license risk management, and regulations are urgently required to prevent the chaos of license proliferation.}

        \stepcounter{discussionCount}
        \noindent \textbf{Challenge \arabic{discussionCount}: Missing Fine-Grained Meta-Model of Software Licenses.}
            As a gradual common practice to clarify license restrictions, {license terms are increasingly being formally modeled and operationalized for settling violations as the supply of license knowledge~\cite{lidetector, LiResolver, liu2024liscopelens}.} 
            However, there are still disputes on the meta-model to guide the understanding of software licenses.
        \begin{table}[]
\small
\centering
\caption{Overview of License Identification Methods Used in License Risk Assessment Research}
\label{tab:identificationAnalasis}

\resizebox{0.49\textwidth}{!}{

\rowcolors{1}{white}{gray!20}
\scalebox{0.85}{
\begin{tabular}{l|l|c|c|c|c|c|c|c}
\Xhline{0.8px}
\hiderowcolors
\multirow{2}{*}{\textbf{Tool or First Author}} & \multirow{2}{*}{\textbf{\boyuan{Year}}} & \multicolumn{6}{c|}{\textbf{Name Identification}}          & \multicolumn{1}{c}{\multirow{2}{*}{\textbf{\begin{tabular}[c]{@{}c@{}}Term\\ Identification\end{tabular}}}} \\\cline{3-8}
        &        & \multicolumn{1}{c}{Manual} & \multicolumn{1}{c}{Ninka} & \multicolumn{1}{c}{FOSSology} & \multicolumn{1}{c}{ScanCode} & \multicolumn{1}{c}{Metadata} & \multicolumn{1}{c|}{Others} & \multicolumn{1}{c}{}                \\ \hline
\showrowcolors

German et al.~\cite{codeSiblings}                   & 2009   &    &   & \textcolor[HTML]{00A551}{\checkmark}             &      &      &    &             \\

Di Penta et al.~\cite{di2009source} & 2009   &   &   &       &      & \textcolor[HTML]{00A551}{\checkmark}            &    &             \\

LChecker~\cite{LChecker} & 2010   &   &   &       &      & \textcolor[HTML]{00A551}{\checkmark}            &    &             \\
German et al.~\cite{auditingLicesning}                   & 2010   &    & \textcolor[HTML]{00A551}{\checkmark}         &       &      & \textcolor[HTML]{00A551}{\checkmark}     &              &             \\

Di Penta et al.~\cite{di2010exploratory}& 2010   &   &   & \textcolor[HTML]{00A551}{\checkmark}      &      &             &    &             \\



BAT~\cite{BAT}    & 2011   &    &   &       &      &      & \textcolor[HTML]{00A551}{\checkmark} &             \\

Kenen~\cite{Kenen}   & 2012   & \textcolor[HTML]{00A551}{\checkmark}          & \textcolor[HTML]{00A551}{\checkmark}         &       &      &      &    &             \\



Alspaugh et al.~\cite{heterogeneous2012Alspaugh}& 2012   &   &   &       &      &             &    & \textcolor[HTML]{00A551}{\checkmark}            \\

Mathur et al.~\cite{GoogleViolation}& 2012   &   &   &       &      & \textcolor[HTML]{00A551}{\checkmark}            &    &             \\

CBDG~\cite{CBDG}    & 2014   &    & \textcolor[HTML]{00A551}{\checkmark}         &       &      &      &    &             \\

Wu et al.~\cite{inconsistenciesInLarge-scaleProjects}                       & 2015   &    & \textcolor[HTML]{00A551}{\checkmark}         &       &      &      &    &             \\

Vendome et al.~\cite{vendome2015licensechange}& 2015   &   &\textcolor[HTML]{00A551}{\checkmark}   &       &      & \textcolor[HTML]{00A551}{\checkmark}            &    &             \\





Mlouki et al.~\cite{androidViolation}                   & 2016   &    & \textcolor[HTML]{00A551}{\checkmark}         &       &      &      & \textcolor[HTML]{00A551}{\checkmark} &             \\

SPDX-VT~\cite{SPDXcompatibility} & 2017   & \textcolor[HTML]{00A551}{\checkmark}          &   & \textcolor[HTML]{00A551}{\checkmark}             &      &      &    &             \\

Wu et al.~\cite{inconsistencyInLargeCollections}                & 2017   &    & \textcolor[HTML]{00A551}{\checkmark}         &       &      &      &    &             \\

OSSPolice~\cite{OSSPolice}                       & 2017   &    &   & \textcolor[HTML]{00A551}{\checkmark}             &      & \textcolor[HTML]{00A551}{\checkmark}            &    &             \\

Vendome et al.~\cite{vendome2017licensechange}& 2017   &   &\textcolor[HTML]{00A551}{\checkmark}   &       &      &             &    &             \\

An et al.~\cite{ToxicAndroid}                & 2017   & \textcolor[HTML]{00A551}{\checkmark}          &   &       &      &      &    &             \\

Lotter et al.~\cite{stackoverflowJava}& 2018   &   &  &       &      &\textcolor[HTML]{00A551}{\checkmark}              &    &             \\

Feng et al.~\cite{BinaryViolation}            & 2019   & \textcolor[HTML]{00A551}{\checkmark}          &   &       &      &      & \textcolor[HTML]{00A551}{\checkmark} &             \\ 

Ragk. et al.~\cite{ToxicOnStackOverflow}          & 2019   & \textcolor[HTML]{00A551}{\checkmark}          &   &       &      &      &    &             \\

Baltes et al.~\cite{ToxicGitHub}            & 2019   &    &   &       &      &      & \textcolor[HTML]{00A551}{\checkmark}               &             \\

OSLDetector~\cite{OSLDetector}& 2020   &   &  &       &      & \textcolor[HTML]{00A551}{\checkmark}             &    &             \\

Golubev et al.~\cite{GitHubViolation}                  & 2020   &    & \textcolor[HTML]{00A551}{\checkmark}         &       &      & \textcolor[HTML]{00A551}{\checkmark}            &    &             \\

Moraes et al.~\cite{multi-licensing_JavaScript_ecosystem}& 2021   &   &  &       & \textcolor[HTML]{00A551}{\checkmark}     &              &    &             \\

Serafini et al.~\cite{serafini2022efficient}& 2022   &   &  &       &      &              & \textcolor[HTML]{00A551}{\checkmark}   &             \\

Higashi et al.~\cite{dockerCompatibility}& 2022   &   &\textcolor[HTML]{00A551}{\checkmark}  &       &      &              &    &             \\

Sorrel~\cite{Sorrel}  & 2021   &    &   &       &      &      & \textcolor[HTML]{00A551}{\checkmark}                     &             \\

Qiu et al.~\cite{qiu2021empirical}  & 2021   &    &   &       &      & \textcolor[HTML]{00A551}{\checkmark}     &                      &             \\

Makari et al.~\cite{makari2022prevalence} & 2022   &    &   &       &      & \textcolor[HTML]{00A551}{\checkmark}     &                      &             \\




LiDetector~\cite{lidetector}                      & 2023   &    & \textcolor[HTML]{00A551}{\checkmark}         &       &      & \textcolor[HTML]{00A551}{\checkmark}            &    & \textcolor[HTML]{00A551}{\checkmark}                   \\
LiResolver~\cite{LiResolver}                      & 2023   &    & \textcolor[HTML]{00A551}{\checkmark}         &       &      & \textcolor[HTML]{00A551}{\checkmark}            &    & \textcolor[HTML]{00A551}{\checkmark}                   \\

LicenseRec~\cite{LicenseRec}    & 2023   &    &   &       &  \textcolor[HTML]{00A551}{\checkmark}    & \textcolor[HTML]{00A551}{\checkmark}     &       &              \\

DIKE~\cite{DIKE}    & 2023   &    &   &       &      &      & \textcolor[HTML]{00A551}{\checkmark}                       & \textcolor[HTML]{00A551}{\checkmark}                   \\

SILENCE~\cite{pypi} & 2023   &    &   &       & \textcolor[HTML]{00A551}{\checkmark}            & \textcolor[HTML]{00A551}{\checkmark}            &    &         \\

CODEIPPROMPT~\cite{CODEIPPROMPT}& 2023   &   &  &       &      &\textcolor[HTML]{00A551}{\checkmark}              &    &             \\

Reid et al.~\cite{reid2023applying}& 2023   &   &  &       &      &\textcolor[HTML]{00A551}{\checkmark}              &    &             \\

Harutyunyan et al.~\cite{harutyunyan2023open}& 2023   &   &  &       &      &              & \textcolor[HTML]{00A551}{\checkmark}   &             \\

Liu et al.~\cite{catchTheButterfly}                      & 2024   &    &   &       &      & \textcolor[HTML]{00A551}{\checkmark}            &    &  \textcolor[HTML]{00A551}{\checkmark}            \\





Wu et al.~\cite{baolingfenglicense}& 2024   &   &  &       &      &              & \textcolor[HTML]{00A551}{\checkmark}   &             \\

Antelmi et al.~\cite{antelmi2024analyzing}& 2024   &   &  &       &  \textcolor[HTML]{00A551}{\checkmark}     &              &   &             \\


Sun et al.~\cite{sun2024knowledge}& 2024   &   &  &       &      &  \textcolor[HTML]{00A551}{\checkmark}             &   &             \\

ModelGo~\cite{duan2024modelgo}& 2024   &   &  &       &      &              &   & \textcolor[HTML]{00A551}{\checkmark}             \\

\hline
\hiderowcolors
\multicolumn{2}{c|}{\textbf{Total (43)}}  & 5 & 12 & 4 & 4 & 19 & 9  & 6\\

\Xhline{0.8px}
\end{tabular}
}
}

\end{table}
        
        We reviewed the license identification approaches adopted and mentioned in license risk assessment research to inspect the meta-model behind the understanding of software licenses. 
        As presented in~\Cref{tab:identificationAnalasis}, apart from manual analysis and package metadata, existing research heavily relies on Ninka, FOSSology, and ScanCode to support the identification of licenses. Specifically, there is a clear shift in the analyzed targets in research on license risk detection. Before 2017, these research works mainly relied on these tools to retrieve information, identify software licenses, and then conduct studies at the license level, while after 2023, researchers turned to detecting the detailed license terms and conducting finer-grained analysis on term-level conflicts. This shifting indicates the need for a finer-grained interpretation of software licenses for follow-up research, as well as real-world requirements.

        Current research works based on license terms are gradually sorting out a standardized meta-model for license terms.
        Existing software license platforms, such as TLDRLegal~\cite{tldrlegal}, Choosealicense~\cite{choosealicense}, and OpenEuler~\cite{LicenseShowRoom}, are proposed to index popular software licenses with their terms.
        {Based on these term templates, researchers have also increasingly adopted SOTA techniques to extract license terms~\cite{keclausebench}.}
        However, existing term-based license meta-models still suffer from issues such as low coverage, lack of standardization, and practical limitations. As revealed by Liu et al.~\cite{catchTheButterfly}, there are significant gaps among these license indexing platforms. First, the interpretation of software licenses requires domain knowledge of laws and regulations from experts, which is difficult to automate and extend to cover all software licenses. Next, the interpretation model of software licenses, i.e., term definitions, attitude judgment, conflict determination, also vary among different platforms, which could bias the judgment of users and compromise the real-world adoption of software license tools developed from software engineering disciplines.

        \stepcounter{discussionCount}
        \noindent \textbf{Challenge \arabic{discussionCount}: Limited Investigation of License Compliance.}
        Researchers are increasingly focusing on license compliance to mitigate potential legitimate risks. However, compared to the various obligations, 
        existing concerns are mainly on limited scopes.
        As presented in~\Cref{tab:complianceAnalysis},
        we summarized obligations discussed and emphasized in existing research works on license compliance assessment.
        {Our investigation reveals that most existing research concentrates on obligations of \textit{Relicense}, and some others involve obligations of \textit{Include Copyright}, \textit{Include License}, \textit{Include Notice}, \textit{Disclose Source}, \textit{Sublicense} and \textit{State Changes}. 
        Regarding the terms concluded in previous works~\cite{lidetector, catchTheButterfly}, there are far more obligations to investigate for compliance analysis. 
        {To date, we have not found any research specifically addressing the detection of compliance behaviors related to certain terms in the existing model~\cite{lidetector}, such as \textit{Include Original}, \textit{Rename}, and \textit{Contact Author}. While the risk levels associated with these terms may be lower than those previously mentioned, they still present potential threats, especially given the lack of adequate techniques to detect these compliance violations.}
        
        \begin{table}[]
\fontsize{11.5pt}{12pt}\selectfont
\centering

\caption{Overview of License Obligations Detected in Compliance Assessment Research}
\label{tab:complianceAnalysis}

\resizebox{0.49\textwidth}{!}{

\rowcolors{2}{gray!20}{white} 
\scalebox{0.52}{
\begin{tabular}{l|c|c|c|c|c|c|c|c}
\Xhline{0.8px}
\textbf{\begin{tabular}[l]{@{}l@{}}Tool or First Author\end{tabular}} & \textbf{Year} & 
\textbf{\begin{tabular}[c]{@{}c@{}}Include\\Copyright\end{tabular}} & 
\textbf{\begin{tabular}[c]{@{}c@{}}Include\\License\end{tabular}} & 
\textbf{\begin{tabular}[c]{@{}c@{}}Include\\Notice\end{tabular}} & 
\textbf{\begin{tabular}[c]{@{}c@{}}Disclose\\Source\end{tabular}} &

\textbf{Sublicense} & 
\textbf{Relicense}&
\textbf{\begin{tabular}[c]{@{}c@{}}State\\Changes\end{tabular}}
\\ 
\hline

Di Penta et al.~\cite{di2009source}& 2009 &\textcolor[HTML]{00A551}{\checkmark}&&&&&&\\

Molina et al.~\cite{molina2007fast}& 2010 &&&&\textcolor[HTML]{00A551}{\checkmark}&&&\\

Di Penta et al.~\cite{di2010exploratory}& 2010 &&&&&&\textcolor[HTML]{00A551}{\checkmark}&\\

BAT~\cite{BAT}& 2011 &&&&\textcolor[HTML]{00A551}{\checkmark}&&&\\

Wu et al.~\cite{inconsistenciesInLarge-scaleProjects} & 2015 &&&&&& \textcolor[HTML]{00A551}{\checkmark}&\\

Vendome et al.~\cite{vendome2015licensechange}& 2015 & &&&&&\textcolor[HTML]{00A551}{\checkmark}&\\


Mlouki et al.~\cite{androidViolation} & 2016 &&&&& \textcolor[HTML]{00A551}{\checkmark} & \textcolor[HTML]{00A551}{\checkmark}&\\

Wu et al.~\cite{inconsistencyInLargeCollections} & 2017 &&&&&& \textcolor[HTML]{00A551}{\checkmark}&\\

OSSPolice~\cite{OSSPolice}   & 2017 &  &  &  & \textcolor[HTML]{00A551}{\checkmark} & & &\\

An et al.~\cite{ToxicAndroid}   & 2017 & \textcolor[HTML]{00A551}{\checkmark} & \textcolor[HTML]{00A551}{\checkmark} & \textcolor[HTML]{00A551}{\checkmark} & & & \textcolor[HTML]{00A551}{\checkmark}&\\

Vendome et al.~\cite{vendome2017licensechange}& 2017 & &  &  && &\textcolor[HTML]{00A551}{\checkmark} &\\

Lotter et al.~\cite{stackoverflowJava}& 2018 &  & \textcolor[HTML]{00A551}{\checkmark}  & \textcolor[HTML]{00A551}{\checkmark}  & & &\textcolor[HTML]{00A551}{\checkmark}  &\\

Baltes et al.~\cite{ToxicGitHub}   & 2019 &  &  & \textcolor[HTML]{00A551}{\checkmark} & & & \textcolor[HTML]{00A551}{\checkmark} &\\  

Ragk. et al.~\cite{ToxicOnStackOverflow}   & 2019 &  & \textcolor[HTML]{00A551}{\checkmark} &  & & & &\\ 

Feng et al.~\cite{BinaryViolation}   & 2019 &  &  &  & \textcolor[HTML]{00A551}{\checkmark}& & &\\

Golubev et al.\cite{GitHubViolation}   & 2020 &  &  & & & &\textcolor[HTML]{00A551}{\checkmark}&\\

OSLDetector~\cite{OSLDetector}& 2020 &&&&\textcolor[HTML]{00A551}{\checkmark}&&&\\


CODEIPPROMPT~\cite{CODEIPPROMPT}& 2023 &&\textcolor[HTML]{00A551}{\checkmark}&\textcolor[HTML]{00A551}{\checkmark}&&&&\\

Reid et al.~\cite{reid2023applying}& 2023 &\textcolor[HTML]{00A551}{\checkmark}&\textcolor[HTML]{00A551}{\checkmark}&&&&&\\


ModelGo~\cite{duan2024modelgo}& 2024 &&\textcolor[HTML]{00A551}{\checkmark}&\textcolor[HTML]{00A551}{\checkmark}&\textcolor[HTML]{00A551}{\checkmark}&\textcolor[HTML]{00A551}{\checkmark}&\textcolor[HTML]{00A551}{\checkmark}&\textcolor[HTML]{00A551}{\checkmark}\\

Kapitsaki et al.~\cite{kapitsaki2024generative}& 2024 &&&\textcolor[HTML]{00A551}{\checkmark}&&&&\\

\hline
\hiderowcolors
\multicolumn{2}{c|}{\textbf{Total (21)}}  & 3  & 6  & 6  & 6 & 2 & 11 &1\\

\Xhline{0.8px}
\end{tabular}
}

}
\end{table}

        \stepcounter{discussionCount}
        \noindent \textbf{Challenge \arabic{discussionCount}: Ambiguity Definition in License Context.}
        {The definition of software licenses often involves specific terms from the discipline of law, while some terms, from the perspective of measurable metrics, are still ambiguous to detect, complicating how practitioners understand, utilize, and manage their licenses. However, the academic community has yet to engage in a thorough discussion of these issues.}

        A typical example of ambiguous terms from licenses is the concept of \textit{Derivative Works}.
        Many licenses stipulate that the distribution or modification of source code or derivative works incur specific obligations. 
        Yet, the evaluation of what constitutes a derivative work often lacks a consistent, quantifiable approach. Existing researches in license risk analysis mostly utilize code clone detection tools (e.g., CCFinderX~\cite{CCFinder}, SourcererCC~\cite{SourcererCC}) to identify derived code from third parties, which usually rely on empirical thresholds, such as length of similar codes snippets{~\cite{SourcererCC}} and similarity on features of code semantics (e.g., Abstract Syntax Tree~\cite{astclone2}, Program Dependency Graph~\cite{atvhunter}) to approach the behavior of real copies. However, similar code snippets are not a reliable indicator of actual code reuse, and there remains no clear definition of how such similarities relate to \textit{Derivative Works}. 
        {Although this term is primarily a legal issue needing a clear definition}, these intricacies make automated code derivation determination infeasible, highlighting the need for more robust, objective methodologies in licensing compliance analysis.

        Besides the \textit{derivative works}, some other terms are also not captured by existing tools. For instance, there is no comprehensive taxonomy on the approaches and formats of third-party usage. In early licenses, there are no explanations of the approaches and formats of how artifacts can be used. Afterward, software licenses gradually distinguish artifact formats, such as source code, binaries, and executables, as well as how they are integrated, such as package manager dependencies, static and dynamic links{~\cite{lgpl}}, web API calls~\cite{googleMapLicense}, Software Development Kits (SDKs){~\cite{sdkLicenseAgreement}}, to grant different rights and obligations. However, current research also only focuses on limited types of third-party artifact reuses, and there is a conspicuous dearth of research that systematically categorizes the myriad software distribution methods specified by licenses and undertakes targeted compliance assessments. 
        The community is also gradually introducing new scenarios of software usage that could challenge existing software license systems, such as the rising concern about whether open-source codes could be used to train CodeLLMs~\cite{CODEIPPROMPT}, is not scoped by most of the existing popular licenses. 
        
        {Moreover, some rights and obligations are restricted with specific conditions that are based on 
        abstract, context-dependent, or subjective criteria, making it difficult to interpret the semantics and detect the potential license risks. 
        For example, the BSD 3-Clause No Nuclear License~\cite{BsdNoNuclear} prohibits the use of software in nuclear facilities, while the JSON License~\cite{jsonlicense} includes a moral clause stating that ``The software shall be used for good, not evil." Such terms lack precise definitions, leading to uncertainty in compliance assessment and legal interpretation.}

        These gaps underscore the need for more structured and well-defined models and explanations for license interpretation, along with standardized regulations for more comprehensive and unambiguous license risk management tools.


        \stepcounter{discussionCount}
        \noindent \textbf{Challenge \arabic{discussionCount}: License Tampering During Code Reuse.} 
        With the prevalence of source code reuse in modern software development, license missing or even tampering during user copy-and-paste reuse
        are gradually common. This often involves modifying or completely omitting the original license information, such as removing or omitting header comments that contain copyright notices, 
        which could largely bias the existing solutions on license risk detection.  
        For instance, a developer might copy a function from a project licensed under GPL and integrate it into a proprietary project without disclosing the source or adhering to the GPL’s terms about open-sourcing the \textit{derivative work}. 
        If this project is reused by downstream users, the lack of declarations may prevent text-based license identification tools from detecting GPL content in third-party dependencies, posing a risk of GPL violations.

        The reasons behind such license tampering could vary. Developers could deliberately attempt to avoid the complications of license compliance, or might not understand the implications of mixing code under different licenses, such as the legal requirement to release proprietary software under a GPL if it includes GPL-licensed code. However, legally, this can lead to lawsuits and demands for statutory damages. From a business perspective, companies can suffer severe reputational damage if they are seen as copyright violators, potentially leading to lost business opportunities.

        Moreover, after decades of unlimited source code reuse, there could be already plenty of source codes not associated with their original licenses. Especially, the wide adoption of LLM-based code generation in modern software development could even aggravate the chaos of license tampering. Considering most existing license risk management tools, it is daunting and challenging to properly manage and govern the ecosystem-wide license risks in this context by existing techniques. To this end, comprehensive identification and source tracing of source code reuse is urgently required to clear out the original licenses, especially for those commonly reused code snippets.

         \begin{table}[]
\rowcolors{2}{white}{gray!20}
\centering
\caption{License Risk Remediation Strategies}
\label{tab:remediationTab}

\resizebox{0.49\textwidth}{!}{

\begin{tabular}{l|c|c|c|c|c|c|c|c
}
\Xhline{0.8px}
\hiderowcolors
   \multirow{7}{*}{\makecell[l]{\textbf{Tool or}\\\textbf{First Author}}} 
   & \multicolumn{1}{c|}{\textbf{\makecell[c]{Obligation\\Complement}}}
  & \multicolumn{3}{c|}{\textbf{\makecell[c]{License Altering}}} 
  & \multicolumn{4}{c}{\textbf{\makecell[c]{Component Replacing}}} 
  \\
  \cline{2-9}
   & \makecell[c]{\\[-6em] -}
  & \rotatebox{90}{\textbf{Relicense}} & \rotatebox{90}{\textbf{\makecell[c]{Customization}}} & \rotatebox{90}{\textbf{Negotiation}} & \rotatebox{90}{\textbf{\makecell[c]{Code\\Refactoring}}} & \rotatebox{90}{\textbf{\makecell[c]{Reuse via API}}} & \rotatebox{90}{\textbf{\makecell[c]{Components\\ Replacement}}} & \rotatebox{90}{\textbf{\makecell[c]{Components\\Removal}}} \\ 
  \hline
\showrowcolors

Hammouda et al.~\cite{hammouda2010open} &  &\textbullet   &   &   &   &   &   & \textbullet \\

German et al. \cite{intetrationPatterns} &  & \textbullet  & \textbullet  & \textbullet  &   & \textbullet  & \textbullet  &  \\

Scacchi et al.~\cite{scacchi2012understanding} &  &   &   &   &   & \textbullet  & \textbullet  &  \\

Vendome et al.~\cite{vendome2016assisting} & &\textbullet   &   &   &   &   &   & \\
 
Mlouki et al. \cite{androidViolation} & & \textbullet  &   &   &   &   &   & \textbullet \\

Viseur~\cite{selectionofCloudComputing} & &\textbullet   &   &   &   &   &   & \\

ALP~\cite{ALP}&  &\textbullet   &   &   &   &   &   &\\

findOSSLicense~\cite{FindossLicense2} & &\textcolor[HTML]{00A551}{\checkmark}   &   &   &   &   &   & \\

Sorrel~\cite{Sorrel} & &\textcolor[HTML]{00A551}{\checkmark}   &   &   &   &   &   & \\

 LicenseRec~\cite{LicenseRec} & &\textcolor[HTML]{00A551}{\checkmark}   &   &   &   &   &   &\\

DIKE \cite{DIKE}&  & \textcolor[HTML]{00A551}{\checkmark} &  &  & \textbullet &  &  & \textbullet\\

LiResolver \cite{LiResolver}&  & \textcolor[HTML]{00A551}{\checkmark}  & \textcolor[HTML]{00A551}{\checkmark}  & \textbullet &  &  &  & \\

SILENCE \cite{pypi}&  & \textcolor[HTML]{00A551}{\checkmark} &  & \textbullet &  &  & \textcolor[HTML]{00A551}{\checkmark} & \textcolor[HTML]{00A551}{\checkmark}\\

LiVo \cite{huang2024your} &\textcolor[HTML]{00A551}{\checkmark}& &  &  &  &  & &  \\

\Xhline{0.8px}

\end{tabular}
}
\medskip
\begin{minipage} {0.98\linewidth}
\footnotesize{
\textcolor[HTML]{00A551}{\checkmark}: Strategy was implemented. 
\textbullet\;\,: Strategy was collected or mentioned.}
\end{minipage}
\end{table}   
        \stepcounter{discussionCount}
        \noindent \textbf{Challenge \arabic{discussionCount}: Limited Practicability of License Risk Remediation.}
        Considering the complexity and prevalence of license risks, researchers from both academia and the industry have proposed various solutions to assist and automate the mitigation of potential legitimate issues.
        {Apart from satisfying required conditions, e.g., making up necessary notices and statements~\cite{huang2024your},} 
        typically, existing solutions can be classified into two major types, 1) altering licenses of user projects, such as relicense, license customization, license exception, and negotiation, and 2) replacing third-party components, such as code refactoring, reuse via APIs, component replacement, and component removal, to avoid potential license incompatibility or non-compliance, as presented in~\Cref{tab:remediationTab}. However, these mitigation strategies are still limited by their practicability.

        For license-altering solutions, although there are hundreds of mainstream OSS licenses, there is not always a suitable one to handle all potential legitimate risks. 
        Next, replacing licenses could be challenging to accomplish. An open-source project often involves many contributors, while the license replacement should be granted by all contributors, which could be difficult.
        A typical example is the multi-year efforts of relicensing agreement collection of LLVM~\cite{llvm} to provide patent protection for LLVM users. As for license customization and exceptions, although these solutions could temporarily 
        avoid potential legitimate risks, they could exacerbate the license proliferation in real-world applications, which may lead to a more severe situation for license identification and governance. Moreover, since licenses are usually written in professional languages, freely modifying license text or generating licenses could face potential threats to the rigorousness and legality of generated text. As for negotiation with owners of conflicted components proposed by existing works~\cite{LiResolver, intetrationPatterns}, it is a case-by-case solution and is still not systematically studied yet. 
        
        For third-party component replacement solutions, the primary concern is their impact on the functionality of projects. It's relatively rare for licenses to change during software version updates, thus a common approach is to replace the third-party artifacts with restrictive licenses to those under more permissive ones. However, suitable alternative libraries that meet specific user requirements are not always available. Even when alternatives exist, automating the migration of libraries remains a significant challenge. With the rise of generative software engineering, there is potential for large language models (LLMs) to facilitate this process. Nonetheless, the use of LLMs introduces its own complexities, particularly concerning the copyright status of code generated by LLMs. This indicates that while LLMs could streamline library migration, there remains a considerable path ahead in fully realizing their potential and addressing associated legal concerns.

        \stepcounter{discussionCount}
        \begin{table}[]
\rowcolors{1}{white}{gray!20}
\centering
\caption{license recommendation and selection factors}
\label{tab:recommendationTab}

\resizebox{0.49\textwidth}{!}{

\begin{tabular}{l|c|c|c|c|c}
\toprule

\textbf{Tool or First Author} & 
\textbf{\begin{tabular}[c]{@{}c@{}}Commercial\\Impact\end{tabular}} &
\textbf{\begin{tabular}[c]{@{}c@{}}Community\\Engagement\end{tabular}}&
\textbf{Compatibility} & \textbf{\begin{tabular}[c]{@{}c@{}}Project\\Nature\end{tabular}} & \textbf{\begin{tabular}[c]{@{}c@{}}Developers'\\Requirement\end{tabular}}
\\ \hline

Colazo et al.~\cite{colazo2005copyleft}&\textbullet&\textbullet&\textbullet&&\\
 Stewart et al.~\cite{stewart2005preliminary}&\textbullet&\textbullet&&&\\
Colazo et al.~\cite{colazo2009impact}&\textbullet&\textbullet&&&\\
Lindman et al.~\cite{lindman2011matching}&&\textbullet&\textbullet&&\\
Kashima et al.~\cite{kashima2011investigation}&\textbullet&\textbullet&\textbullet&\textbullet&\\
Ghapanchi et al.~\cite{ghapanchi2011impact}&&\textbullet&&&\\
Vendome et al.~\cite{vendome2015licensechange}&\textbullet&\textbullet&\textbullet&&\\
Vendome et al.~\cite{developerRef2}&\textbullet&\textbullet&\textbullet&&\\
Vendome et al.~\cite{vendome2017licensechange}&\textbullet&\textbullet&\textbullet&&\\
Medappa et al.~\cite{medappa2017license}&\textbullet&&&&\textbullet\\
Gamalielsson et al.~\cite{gamalielsson2017licensing}&&\textbullet&\textbullet&&\\

ALP~\cite{ALP}&& & \textcolor[HTML]{00A551}{\checkmark} & \textcolor[HTML]{00A551}{\checkmark}   &   \\

findOSSLicense~\cite{FindossLicense2}&&  & \textcolor[HTML]{00A551}{\checkmark}  & \textcolor[HTML]{00A551}{\checkmark}  & \textcolor[HTML]{00A551}{\checkmark} \\

LicenseRec~\cite{LicenseRec}&& & \textcolor[HTML]{00A551}{\checkmark}    &   & \textcolor[HTML]{00A551}{\checkmark}\\

Zhang et al.~\cite{recommendationPower}&& &   & \textcolor[HTML]{00A551}{\checkmark}  &   \\

\hiderowcolors

\bottomrule
\end{tabular}
}
\medskip
\begin{minipage} {0.98\linewidth}
\footnotesize{
\textcolor[HTML]{00A551}{\checkmark}: Factor was used for implementation. 
\textbullet\;\,: Factor was mentioned.}

\end{minipage}
\end{table}

        \noindent \textbf{Challenge \arabic{discussionCount}: Inadequate Requirement Mining for License Recommendation.}
        {Besides remediating license risks, license recommendation techniques also play important roles in avoiding potential legitimate risks. Existing works have identified various factors influencing license selection and have developed recommendation tools to assist developers, as identified in~\Cref{tab:recommendationTab}, while these existing tools fail to integrate these considerations in subtle ways.}


        {
        For instance,
        one of the primary 
        concerned factors
        involves compatibility checks, akin to those used in license risk remediation. This approach, however, struggles with a limited availability of compatible licenses, often leading to recommendations that do not fully meet the specific needs of different projects. As a result, developers and organizations are increasingly turning towards customized licenses, resulting in license proliferation, which could create a fragmented landscape of licensing options that may not always adhere to best practices for ecosystem-wide license management and standardization.
        Apart from this, requirements from developers and project natures are also considered by existing methods. However, the gathering of developer requirements is confined largely to the terms specified within licenses, failing to capture the broader, more detailed requirements of the application scenarios for which the software is intended, {such as moral reasons and community influence~\cite{developerRef2}}.
        These tools also fail to account for the complexities of \textit{commercial impact} and \textit{community engagement}, such as whether a recommended license aligns with corporate preferences to attract sponsorship, guarantees openness, or fosters community involvement in development and maintenance~\cite{stewart2005preliminary,colazo2005copyleft,medappa2017license}.}

        {Furthermore, while research has outlined numerous factors affecting license selection, a systematic discussion is still lacking. Existing research has yet to establish a comprehensive framework that fully summarizes and organizes all considerations. There is no clear consensus on how to weigh or balance different factors. As a result, current approaches remain fragmented, limiting their effectiveness in providing well-rounded recommendations for developers in diverse contexts.}

    \subsection{Opportunities for Future Research}
    \setcounter{discussionCount}{0}

    Considering the shared trends among the challenges currently faced in License Risk Management and the progression of existing licenses, we identify potential opportunities for future advancements in this research area.

    \stepcounter{discussionCount}
    \noindent \textbf{Opportunity \arabic{discussionCount}: Standardized License Meta-Model and Framework.} 
    As revealed by the SLR, the lack of standardization becomes increasingly problematic as the proliferation of licenses grows, leading to confusion and complexity for management. As user requirements evolve and software reuse scenarios increase in complexity, the need for standardized definitions and terms becomes even more critical. Standardizing these elements would improve understanding of the licenses and enhance risk management associated with them. There is a promising potential for experts to endorse the relationships between existing licenses and these standardized definitional entities, progressively advocating for the incorporation of more granular license content as the standard for future licensing practices.

    {However, existing license knowledge bases mainly focus on prevalent licenses, presenting significant deficiencies and inadequacies for the modeling of license semantics, especially for conditions and restrictions raised from the emerging requirements of developers, such as the evolving ways of reuse, new types of obligations, and exceptions for corner cases, etc.}
    {
    For example, to address the limitations of traditional licenses in expressing new ways of reusing open source LLMs, such as leveraging a model's weights, parameters, activations, and other related materials for methods like fine-tuning, evaluation, or distillation to improve other models, many large model providers carefully design customized licenses during open-source releases to avoid ambiguity~\cite{llamaLicense, deepseekLicense}.}
        
    {To this end, a comprehensive analysis and standardization of software licenses could not only be useful for governance but also inspire the flexibility to satisfy the demands of real-world license applications efficiently.}    
    {However, constructing such a meta-model is a highly challenging task that requires substantial collaborative efforts from the OSS community, industry, and academia, especially from the legal field to ensure professional and precise interpretations of license semantics and risks. Quantifying such details will support more accurate license management and help reduce license content disputes.}

    \stepcounter{discussionCount}
    \noindent \textbf{Opportunity \arabic{discussionCount}: Measurable Metrics for Automated License Violation Detection.} 
    Current research on license risk detection in software management has predominantly focused on the more observable aspects such as the inclusion of licenses, copyrights, {notices}, and conflicts arising from contradictory licensing terms.
    {However, since these compliance requirements cannot be effectively measured, automated detection is infeasible, making manual review indispensable.}
    {Furthermore, many obligations within the existing license term model remain undetectable in compliance. For example, proper use of trademarks, which requires software to avoid misleading representations that could confuse the trademark with a generic term, is rarely monitored effectively. The adherence to patent stipulations and handling of promotional texts, such as avoiding unauthorized endorsements or implied associations, are also critical yet difficult to quantify.}
    The development of measurable metrics, to gauge these less tangible aspects of licensing could revolutionize the approaches to managing and enforcing software licenses.

    Another complex area involves the legal interpretation of terms used in software licenses, which are often steeped in domain-specific jargon that does not easily lend itself to quantifiable metrics. For instance, the term \textit{derivative works} is frequently used in licenses but lacks a clear, universally accepted definition. This ambiguity poses a challenge that determining whether a piece of software that shares similarities with another is simply a coincidence or actually constitutes a \textit{derivative work} can be highly subjective and contentious. Additionally, the scope of what is considered source code similarity can vary, further complicating efforts to automate the detection of such violations. These issues highlight the necessity for more sophisticated tools that can interpret and measure these complex legal concepts within the framework of software license management. Moreover, the interpretation of these domain-specific terms is highly related to local proprietary and copyright laws, different justice systems could result in inconsistent outcomes in similar legal cases, and more investigations on lawsuit cases and corporations with legal professions are expected to strengthen the alignment of software license risk management.

    To address these complexities, there is a pressing need for closer collaboration between the academic field of software management and legal experts specializing in intellectual property and software licensing. By fostering partnerships between these disciplines, it is possible to bridge the gap between legal theory and practical, measurable enforcement in software license management. Such interdisciplinary efforts are crucial for advancing research in license violation detection and ensuring that software management evolves to meet the changing landscapes of law and technology.

    \stepcounter{discussionCount}
    \noindent \textbf{Opportunity \arabic{discussionCount}: License Management of Contributors.}     
    Developing a single OSS project frequently involves numerous contributors, who hold the copyrights to their contributions. However, these contributors could potentially have various views on the roadmap, and management principles of OSS projects, which could lead to divergences. Historical instances, such as the fork of XFree86 into X.Org Server due to license disagreements~\cite{xfree86}, and the creation of LibreOffice in response to dissatisfaction with Oracle's management of the OpenOffice.org suite~\cite{LibreOffice}, exemplify the impact of such divergences. To mitigate disputes and streamline management, agreements such as the Contributor License Agreement (CLA) and the Developer Certificate of Origin (DCO) have been instituted~\cite{WikiCLA, WikiDCO}. These agreements are intended to ensure that contributors retain copyright over their code and agree to transfer these rights to the project maintainers.

    Nonetheless, according to our review, we have found no papers focusing on the implementation and impact of CLAs and DCOs. 
    There are broad directions for future research works to address the critical problem, regarding the prevalence of these agreements in OSS projects. 
    For instance, the effectiveness of CLAs and DCOs in protecting intellectual property rights, the administrative burdens they impose on contributors and maintainers, such as standardized understanding, violation risks, mitigation strategies, and their role in fostering or hindering collaboration within OSS communities. Additionally, the legal implications of these agreements in different jurisdictions and their compatibility with various licensing models represent significant challenges that require detailed examination. 
    To address these issues, future academic endeavors could benefit from collaborations with open-source repositories and developers to investigate the current utilization and evolution of DCOs and CLAs.

    \stepcounter{discussionCount}
    \noindent \textbf{Opportunity \arabic{discussionCount}: License Risks for Generative Software Engineering.} 
    The prevalence of generative software engineering based on LLMs has seen a significant rise in recent years. These models, such as CodeGen~\cite{codegen}, Copilot~\cite{copilot}, and ChatGPT~\cite{ChatGPT}, have demonstrated remarkable capabilities in generating code, automating repetitive tasks, and even assisting in software design and debugging. Developers increasingly rely on these models to streamline workflows and achieve higher efficiency. 
    However, the integration of LLMs in software engineering introduces potential threats and challenges related to license management. One significant concern is the inadvertent incorporation of licensed code snippets generated by LLMs, which could lead to intellectual property violations~\cite{copilotSue, copilotIPdeclare}. Given the vast training data these models are exposed to, it is challenging to ensure that the generated code complies with open-source licensing requirements, {since it is still a gray area 
    to determine whether similar generated code constitutes a derivative of copyrighted work~\cite{xu2024licoeval}.} 
    Additionally, the dynamic nature of LLMs, which continuously learn and adapt, complicates the tracking of license compliance over time. The opacity of these model training datasets further exacerbates this issue~\cite{CODEIPPROMPT}, making it difficult to verify the provenance and licensing status of the generated code.

    To address these challenges, research should be further conducted for license management in the context of generative software engineering. One promising area is the license compliance check for code generated by LLMs and ensuring it adheres to relevant rights and obligations by its original licenses. Additionally, there is a need for transparency in the training datasets of LLMs, allowing for better auditability and verification of code origins. Research could also explore the development of standardized frameworks for integrating license management into the process~\cite{kapitsaki2024generative} of adoption practices of generated code by LLMs, ensuring that licensing considerations are embedded from the outset. 
    {Moreover, another critical issue that may be beyond the scope of OSS license management is that there are urgent needs for a more robust framework of legal explanations of granted rights and obligations in the generative context, such as the boundary of derivative work, rights of use in LLM-specific scenarios, etc. However, this may require further dedication from academia, industry, open-source communities, and laws to work together and derive best practices and guidelines to navigate the complexities of license management in the era of generative software engineering.}
    

    \subsection{Recommendation for Practitioners}

    {In this section, we offer actionable advice for different stakeholders, based on the insights gained from our study.}

    {\textbf{Recommendations for OSS Developers:}}
    
    {1) Effective license management is crucial for OSS developers to ensure legal compliance and foster community trust. As stipulated in existing standards, e.g., ISO/IEC 5230~\cite{iso5230}, software license compliance should be rigorously restricted in the software supply chain throughout software lifecycles. We encourage OSS developers to maintain explicit and up-to-date licensing information in their repositories, i.e.,  a clear LICENSE file at the root of the repository.}
    
    {{2) When selecting software licenses, the considerations outlined in~\Cref{tab:recommendationTab} provide a useful starting point for decision-making. However, existing tools based on these factors may not account for all relevant concerns or usage scenarios. Therefore, developers should still assess additional factors according to the specific context and goals of their projects.}}
    
    {3) Regularly auditing third-party dependencies to verify that their licenses are compatible with the chosen license, by utilizing SCA tools, is also highly recommended. 
    Specifically, it is important to carefully keep an eye on the obligations and conditions to fulfill before integration, and tools at term-level~\cite{lidetector,DIKE,huang2024your} and existing license interpretation websites~\cite{tldrlegal, choosealicense}, can be helpful.}
    
    {4) OSS developers could receive massive commits or PR to their repositories, or they contribute to others' repositories, it is also important to clearly document and review how these contributions are licensed, preferably through a Contributor License Agreement (CLA) or Developer Certificate of Origin (DCO), to avoid future legal complexities.} 
    
    5) Developers should also stay informed about licensing trends and best practices to address any emerging risks or conflicts proactively, helping to safeguard your project's legal integrity and reputation {(e.g., by following initiatives and discussion of OSI and the OpenChain Project~\cite{OPENCHAINACT})}.

    {\textbf{Recommendations for Users:}}
    
    {1) For users of OSS, understanding license management is key to ensuring proper and legal use of the software. Users should always review the license associated with an OSS project before reuse, especially in commercial or proprietary environments, to understand any obligations or restrictions.}
    
    {2) Prevalent licenses like MIT or Apache 2.0 are permissive, allowing broad use with minimal requirements, while copyleft licenses like GPL impose stricter conditions, such as sharing derivative works under the same license. Therefore, it is also important to pay attention to clauses about attribution, redistribution, and modifications to ensure compliance.}
    
    {3) As revealed by existing research~\cite{huang2024your}, less concerned obligations, such as placing appropriate notices, are also easy to overlook. {Users should examine each license obligation systematically based on the existing license term framework~\cite{catchTheButterfly}. When faced with ambiguous definitions, legal professionals should be consulted to ensure accurate understanding.}}
    
    {4) If OSS is integrated into users' own projects, they should carefully verify that the licenses are compatible with the intended use and distribution model.}
    
    {5) When using OSS in production environments, SCA tools are highly encouraged to integrate to help track and manage licenses, especially the rights granted on different integration ways~\cite{wintersgill2024law}, to ensure adherence to all legal requirements and maintain proper documentation.}}
    
    
    {\textbf{Recommendations for Third-Party Auditors:}}
    
    {1) For third-party auditors evaluating OSS projects, a more comprehensive approach to license management is essential. As reviewed in existing research~\cite{nonOsiApprovedLicenses}, auditors could suffer from the limited scope of licenses. They should not only focus on popular licenses like MIT and GPL, but also include more niche licenses {to address existing license proliferation}.}
    
    {2) Accurate license identification should also extend beyond the LICENSE file to include code comments, sub-folders, and documentation, uncovering hidden or conflicting licenses. 
    Based on these, incompatibility and noncompliance detection should consider both license types and the context of usage, like dynamic versus static linking, for precise risk assessment.}
    
    {3) Term-based checks are also crucial, apart from specific license terms such as redistribution and export rights that are directly related to commercial activities, obligations that are easily overlooked, such as including necessary notices and copyright information, should also be included by auditors.}
    
    {4) More practical remediation strategies are highly encouraged to be implemented, such as replacing incompatible dependencies or isolating problematic components, these could significantly help developers and users with flexible paths to achieve compliance without sacrificing functionality.}

\subsection{Limitations and Threats to Validity}\label{sec:threats}

We also discuss the potential threats to the validity and limitations of this SLR as follows.

\noindent $\bullet$ \textbf{Construct Validity.}
{One potential threat arises from the ambiguity and inconsistency of definitions and terminology across existing literature, such as \textit{conflict} and \textit{compatibility}. To address this issue, we clarified key concepts in Section II. Building on this foundation, we developed a unified taxonomy to ensure consistent and systematic classification and synthesis of the literature. In addition, the manual processes in this study could be subjective. First, some screening criteria (ie, EC3 and EC6) were applied entirely based on the author's judgment. Second, the classification of papers also involved potential bias. To mitigate subjectivity, three authors independently conducted manual analyses, cross-validated their results, and resolved any discrepancies through consensus discussions. To further enhance the reliability, we calculated the average Cohen’s $\kappa$ scores for both tasks (i.e., 0.95 and 0.87), indicating a high level of agreement.}

\noindent $\bullet$ \textbf{Internal Validity.}
{Including low-quality or irrelevant literature may threaten the internal validity. We carefully constructed search keywords and queried five prominent electronic databases (IEEE Xplore, ACM Digital Library, SpringerLink, ScienceDirect, and DBLP) to ensure broad coverage of relevant papers. We further established criteria (i.e., EC3 and EC6) to confirm the relevance of selected papers.
In addition, to avoid misleading and unsupported claims, it is necessary to exclude low-quality papers. However, this approach may inadvertently exclude some high-quality work. To address this, we retain only papers published in venues listed in the latest CORE ranking, which is widely recognized as a reliable indicator of venue quality.
Since the dataset we adopted may also miss high-quality papers, we employed an iterative snowballing process to include potentially overlooked works.}

\noindent $\bullet$ \textbf{External Validity.}
{Because there is no established taxonomy for license management, it’s hard to connect academic research to real-world practice. We therefore propose a taxonomy based on the workflow of SCA tools, since these tools are most related to legitimate issues of OSS in the industry. Moreover, as no comprehensive list of SCA tools exists, we followed a widely recognized enterprise report~\cite{synopsys2024report} that summarizes the leading SCA tools to select representative study objectives.}

\section{Conclusion}\label{sec:conclusion}

In this paper, we conducted an SLR on 80 OSS license management-related papers, categorizing the research into three key areas {based on industrial practices}: license identification, risk detection, and risk mitigation. {Based on these, we conducted a comprehensive review to summarize the current state of research, highlight the differences with industry practices, and discuss the challenges, opportunities, and recommendations for different stakeholders.} We hope our review and in-depth discussion can assist both academic researchers and industry practitioners in enhancing the governance of legitimate software risks, thereby improving practices within the software engineering community.

\section*{Acknowledgments}
This work was supported by the National Key Research and Development Project (No. SQ2024YFE0201727).
\chengwei{It is also supported by the National Research Foundation, Singapore, and DSO National Laboratories under the AI Singapore Programme (AISG Award No: AISG2-GC-2023-008); by the National Research Foundation Singapore and the Cyber Security Agency under the National Cybersecurity R\&D Programme (NCRP25-P04-TAICeN).
Any opinions, findings and conclusions, or recommendations expressed in these materials are those of the author(s) and do not reflect the views of National Research Foundation, Singapore, Cyber Security Agency of Singapore, Singapore.}
\balance

\bibliographystyle{IEEEtran}
\bibliography{myref}

\newpage

\end{document}